\documentclass[aps,preprintnumbers,amsmath,amssymb,nofootinbib,eqsecnum,preprintnumbers]{revtex4}

\usepackage{graphicx} % Required for inserting images
\usepackage[a4paper,margin=1.25in]{geometry}
\usepackage{amsmath}
\usepackage{hyperref}
\usepackage{float}
\usepackage{xcolor}
\usepackage{bigints}
\usepackage[utf8]{inputenc}
\usepackage{subcaption}
\usepackage{amsfonts,amssymb,amscd,amsthm,amsfonts,epsfig,enumerate,yfonts}
\usepackage{physics}
\usepackage[font=small,skip=0pt]{caption}
\usepackage{float}
\usepackage{multirow}
\usepackage{makecell} 
\usepackage{tikz}
\usepackage{ragged2e}
\usetikzlibrary{positioning}
\usepackage[font=small,skip=7pt]{caption}
\begin{document}
\title{Shadow of F(R)-Euler-Heisenberg Black Hole and Constraints from EHT Observations}
\author{Khadije Jafarzade$^{1}$}
\email{khadije.jafarzade@gmail.com}
\author{Saira Yasmin$^{2}$}
\email{sairayasmeen555@gmail.com}
\author{Mubasher Jamil$^{2}$}
\email{mjamil@sns.nust.edu.pk}
\affiliation{$^2$School of Natural Sciences, National University of Sciences and Technology (NUST), H-12, Islamabad 44000, Pakistan}
\affiliation{$^1$Department of Theoretical Physics, Faculty of Science, University of Mazandaran, P. O. Box 47416-95447, Babolsar, IRAN}
\date{\today}

\begin{abstract}
\begin{justify}
This study investigates the optical properties of a static, spherically symmetric, electrically charged black hole in $f(R)$ gravity, coupled with Euler-Heisenberg ($\mathit{EH}$) nonlinear electrodynamics (NLED). By analyzing photon trajectories in this background, we demonstrate how the model parameters influence light propagation, leading to broader ranges for both lensed paths and photon rings. We identify parameter regions that allow for physically consistent black hole shadows, characterized by a photon sphere located outside the event horizon and a shadow forming beyond it. These viable regions expand with increasing electric charge and $f_{R_{0}}$, highlighting the interplay between gravitational and electromagnetic effects. Constraining the model using Event Horizon Telescope (EHT) observations of M87*, we find that de Sitter (dS) black holes remain compatible with the data, whereas anti-de Sitter (AdS) solutions are disfavored under low charge and $f_{R_{0}} > -1$. Finally, our analysis of the energy emission rate shows that higher electric charge enhances black hole evaporation, while stronger NLED effects and larger $f_{R_{0}}$ values suppress it.

\end{justify}
\end{abstract}

\maketitle
\section{Introduction}
For many years, the NLED extension of the Reissner-Nordström solution to the Einstein-Maxwell equations has been a topic of considerable interest. One of the most well-known examples is the gravitating Born-Infeld (BI) theory \cite{born1934foundations}. Static, charged black holes in the context of gravitating NLED were initially explored in the 1930s \cite{hoffmann1937choice}. More recently, the connection between string theory and D-brane physics, which leads to Abelian and non-Abelian BI-like Lagrangians in the low-energy limit \cite{tseytlin1997non}, has reignited interest in these nonlinear models. Asymptotically flat, static, spherically symmetric black hole solutions within the Einstein-BI framework have been derived in several studies \cite{demianski1986static}.

\noindent The generalization of exact solutions for spherically symmetric BI black holes with a cosmological constant in arbitrary dimensions has been explored in \cite{fernando2003charged, cai2004born}, as well as in other gravitational backgrounds \cite{aiello2004exact}. Over the past few decades, various models of NLED leading to static and spherically symmetric structures have been investigated. These include theories with nonlinear Lagrangians that are general functions of the gauge invariants (\( F_{\mu\nu} F^{\mu\nu} \) and \( F_{\mu\nu} \hat{F}^{\mu\nu} \) )\cite{diaz2010electrostatic}, theories involving logarithmic functions of the Maxwell invariant \( F_{\mu\nu} F^{\mu\nu} \) \cite{soleng1995charged}, and those with generalized nonlinear Lagrangians \cite{de1994non}, which can give rise to the BI Lagrangian and the weak-field limit of the $\mathit{EH}$ effective Lagrangian \cite{heisenberg1936folgerungen}. The study of static and spherically symmetric black holes coupled to NLED, particularly in the weak-field limit of the $\mathit{EH}$ effective Lagrangian as a low-energy approximation of the BI theory, was presented in \cite{yajima2001black}. There have also been attempts to obtain regular, singularity-free static and spherically symmetric black hole solutions in the context of gravitating NLED \cite{dymnikova2004regular}. Further, generalizations of spherically symmetric black holes in higher dimensions within the theory of nonlinear Lagrangians of power functions of the Maxwell invariant have been explored \cite{mazharimousavi2010theorem}. Lastly, rotating black branes \cite{dehghani2007thermodynamics} and rotating black strings \cite{hendi2010rotating}in the Einstein-BI theory have also been investigated.

\noindent The effective Lagrangian for nonlinear electromagnetic fields was first formulated by Heisenberg and Euler using the Dirac electron-positron theory \cite{heisenberg1936folgerungen}. Schwinger later reformulated this non-perturbative one-loop effective Lagrangian within the framework of quantum electrodynamics (QED) \cite{schwinger1951gauge}. This effective Lagrangian describes the phenomenon of vacuum polarization, where its imaginary part represents the probability of vacuum decay through electron-positron pair production. When electric fields exceed the critical value \( E_c = \frac{m^2 c^3}{e \hbar} \), the energy of the vacuum can be reduced by the spontaneous creation of electron-positron pairs \cite{heisenberg1936folgerungen,schwinger1951gauge}. For many years, both theoretical and experimental physicists have been interested in the electron-positron pair production from the QED vacuum and vacuum polarization induced by an external electromagnetic field \cite{ruffini2010electron}.

The \( F(R) \) gravity theory is a modified gravitational framework where the action is expressed as a general function of the scalar curvature \( R \). This theory has the potential to describe both the early and late cosmological evolution. \( F(R) \) gravity can also account for dark energy and dark matter without introducing new unobserved material, consistent with experimental data \cite{copeland2006dynamics,clifton2012modified}. Modifying the Lagrangian of General Relativity (GR) not only affects the dynamical behavior of the universe but can also influence the dynamics at galactic and solar system scales. Therefore, theories of gravity with higher-order curvature corrections provide a more comprehensive understanding of gravity.

The Einstein-Hilbert action, which reproduces the field equations of GR, is linear in the Ricci scalar \( R \). By altering the action to include nonlinear terms in the Ricci or Riemann curvatures, many viable modified gravitational theories have been proposed by the scientific community to describe the early universe's cosmic evolution. Most of these theories utilize a gravitational Lagrangian containing some of the four possible second-order curvature invariants. Furthermore, many models incorporating higher-order curvature invariants as functions of \( R \) have been introduced, resulting in various \( f(R) \) gravitational models \cite{cognola2008class,vainio2017f,nojiri2024photon}. 

In addition to the potential of these theories to eliminate contributions from curvature invariants other than the Ricci scalar \( R \), they also help address the Ostrogradski instability issue, which is typical of higher-derivative theories \cite{woodard2007avoiding}. The first modification of GR could be traced back to \cite{buchdahl1970non}, where a natural extension of GR involves adding terms like \( R^n \), where \( n \) is a constant, similar to the Starobinsky model \( F(R) = R + \epsilon R^2 \), where \( \epsilon \) is a constant \cite{starobinsky1980new}. For \( n < 0 \), the \( R^n \) term can investigate the late-time universe and describe self-accelerating vacuum solutions \cite{carroll2003can}. However, such solutions are plagued by instabilities \cite{faraoni2005stability} and face strong constraints from solar system tests \cite{chiba20031}. To circumvent these issues, scientists have proposed the \( f(R) \) gravitational theory, which can accommodate a broad range of phenomena. The framework of \( f(R) \) gravity has been applied in various fields, including gravitational wave detection \cite{corda2008primordial}, early-time inflation \cite{bamba2008inflation}, cosmological phases \cite{nashed2018spherically}, the singularity problem stability of solutions, and many other areas \cite{faraoni2005modified}.
Quantum Electrodynamics (QED), beyond being the foundation of modern electromagnetic theory, provides a highly refined framework for describing electromagnetic interactions—one that has been rigorously confirmed through experiments. Recognizing that \( F(R) \) gravity has the potential to account for a range of astrophysical and cosmological phenomena, it becomes important to study how the $\mathit{EH}$ Lagrangian couples to \( F(R) \) gravity. This leads to the formulation of the \( F(R) \)-$\mathit{EH}$ Lagrangian. The $\mathit{EH}$ theory offers a more accurate low-energy approximation of QED than Maxwell’s theory, particularly in regimes involving strong electromagnetic fields \cite{sekhmani2024electrically,nojiri2011unified,nojiri2017modified}.

Within the $\mathit{EH}$ framework, the vacuum behaves like a nonlinear medium, where effects such as polarization and magnetization emerge from virtual charge clouds that give rise to effective physical currents and charges \cite{obukhov2002fresnel}. Owing to the appealing physical features of the $\mathit{EH}$ model, coupling its Lagrangian with the Ricci scalar over the spacetime volume becomes a natural step in the search for black hole solutions.

The first application of nonlinear $\mathit{EH}$ electrodynamics in GR to construct a black hole solution can be traced to \cite{yajima2001black}. Electrically charged black holes were investigated in \cite{ruffini2013einstein}, while the geodesic structure of these solutions was the focus in \cite{amaro2020geodesic}. Furthermore, significant studies have explored various aspects of dyonic and magnetically charged black holes in $\mathit{EH}$ theory, including quasi-normal modes \cite{breton2021birefringence}, thermodynamic properties \cite{dai2023thermodynamic}, and related phenomena. 
The advent of the  EHT marked a pivotal moment in observational astrophysics, providing the first direct images of supermassive black holes at horizon-scale resolution \cite{akiyama2019even}. These observations visualize key relativistic effects, where intense gravitational fields bend light trajectories, creating a distinct dark region, known as the black hole shadow, encircled by a luminous ring of trapped photons \cite{synge1970equations}. Motivated by this, various theoretical frameworks, including those with modifications to gravity or non-linear electromagnetic interactions, have been explored to model black hole shadows in more complex settings \cite{hendi2010rotating,yasmin2025shadow,jafarzade2024kerr,vagnozzi2023horizon,allahyari2020magnetically,nojiri2025black}.  In this study, we focus on this theory to examine the shadow cast by a static, spherically symmetric black hole and assess its compatibility with the EHT image data of M87$^\ast$.

The structure of this paper is organized as follows. In Section~\ref{sec2}, we briefly review the action for the \( F(R)\!-\!\mathit{EH} \) gravity model, where \( F(R) \) gravity is coupled to \(\!\mathit{EH} \)- NLED, and we discuss the corresponding black hole solution in this framework. We then derive the static, spherically symmetric, electrically charged black hole solution in this framework. Section~\ref{sec3} focuses on constructing the effective spacetime metric associated with photon propagation in the electrically charged \( F(R)\!-\!\mathit{EH} \) black hole background. In Section~\ref{sec4}, we analyze the optical properties of this spacetime by deriving the null geodesic equations that govern photon trajectories. Subsection~\ref{sec4a} presents a detailed examination of photon motion confined to the equatorial plane. In Subsection~\ref{sec4b}, we compute the black hole shadow radius and evaluate the ratio between the shadow size and the photon sphere radius to characterize the optical structure. Subsection~\ref{sec4c} compares our theoretical shadow predictions with the observational data from the EHT image of M87$\ast$, placing constraints on the model parameters. In Subsection~\ref{sec4d}, we investigate the black hole's energy emission rate. Finally, in Section~\ref{sec5}, we summarize our findings and present concluding remarks.

 \section{ ACTION AND FIELD EQUATIONS OF F(R)-{EH} THEORY} \label{sec2}
\begin{justify}
 The focus of this section is on the interaction between NLED fields as $\mathit{EH}$ terms within the framework of \( F(R) \) gravity. To properly analyze this, we consider the action, which is given by \cite{sekhmani2024electrically}
\begin{align}
\mathit{I}_{F(R)} = \frac{1}{16\pi} \int_{\mathcal{M}} d^4x \sqrt{g} \left( F(R) - \mathcal{L}(\mathcal{S}, \mathcal{P}) \right).
\end{align}
Here, \( g = \det(g_{\mu \nu}) \) represents the determinant of the metric tensor \( g_{\mu \nu} \), \( F(R) \) is the modified gravitational function in terms of the Ricci scalar \( R \), and \( \mathcal{L}(\mathcal{S}, \mathcal{P}) \) corresponds to the Lagrangian density for the $\mathit{EH}$ term, which is related to the NLED fields.
For the purposes of this analysis, we assume the modified gravitational function to be $
F(R) = R + f(R)$
where \( f(R) \) is an arbitrary function of the Ricci scalar. Additionally, for simplicity in the theoretical discussion, we set the Newtonian gravitational constant and the speed of light to unity: \( G = c = 1 \).
This framework helps us explore the coupling between the $\mathit{EH}$ terms and the \( F(R) \) gravity theory. The $\mathit{EH}$ Lagrangian \( \mathcal{L}(\mathcal{S}, \mathcal{P}) \) is given by \cite{heisenberg1936folgerungen}

\begin{align}
\mathcal{L}(\mathcal{S}, \mathcal{P}) = -\mathcal{S} + \frac{a}{2} \mathcal{S}^2 + \frac{7a}{8} \mathcal{P}^2,
\end{align}
where \( a = \frac{8\alpha^2}{45 m_e^4} \) is the $\mathit{EH}$ parameter, with \( \alpha \) being the fine structure constant and \( m_e \) the electron rest mass. The scalars \( \mathcal{S} \) and \( \mathcal{P} \) are expressed as
\begin{align}
\mathcal{S} = \frac{\mathcal{F}}{4}, \quad \mathcal{P} = \frac{\hat{\mathcal{F}}}{4},
\end{align}
where \( \mathcal{F} = F_{\mu\nu} F^{\mu\nu} \) is the Maxwell invariant, with $F_{\mu\nu}= \partial_\mu A_\nu-\partial_\nu A_\mu$ is the electromagnetic field
 strength of which $A_\mu$ is the gauge potential. Furthermore, the invariant \( \hat{\mathcal{F} }\) can be defined as \( \hat{\mathcal{F}} = F_{\mu\nu} \hat{F}^{\mu\nu} \), where \( \hat{F}_{\mu\nu} = \frac{1}{2} \epsilon_{\mu\nu}  \textcolor{white}{i}^{\rho a}F_{\rho a} \). The condition  \( a = 0 \), vanishes the non-linear electromagnetic behavior of the $\mathit{EH}$ theory, resulting in the linear electromagnetic field of Maxwell's theory ( (\(
\mathcal{L}(\mathcal{S}) = -\mathcal{S}
\)).
In the context of NLED, there are two settings available. The first setting uses the electromagnetic field tensor \( F_{\mu\nu} \), known as the \( \mathcal{F} \)-setting. The second setting is the \( P \)-setting, where the main field \( P_{\mu\nu} \) is defined as
\begin{align}
P_{\mu\nu} = - \left( \mathcal{L}_{\mathcal{S}} F_{\mu\nu} + \hat{ F_{\mu\nu} }\mathcal{L}_{\mathcal{P}} \right),
\end{align}
where \( \mathcal{L}_{\mathcal{S}} = \frac{\partial \mathcal{L}}{\partial \mathcal{S}} \) and \( \mathcal{L}_{\mathcal{P}} = \frac{\partial \mathcal{L}}{\partial \mathcal{P}} \). When NLED is associated with the $\mathit{EH}$ term, \( P_{\mu\nu} \) takes the form
\begin{align}
P_{\mu\nu} = (1 - a F) F_{\mu\nu} - \hat {F_{\mu\nu}} \frac{7a}{4}\mathcal{P}.
\end{align}
The two independent invariants \( P \) and \( O \) associated with the \( P \)-setting are given by
\begin{align}
P = -\frac{1}{4} P_{\mu\nu} P^{\mu\nu}, \quad O = -\frac{1}{4} P_{\mu\nu} \textcolor{white}{i}^\ast P^{\mu\nu},
\end{align}
where $\textcolor{white}{i}^* P_{\mu\nu}=\frac{1}{2\sqrt{-g}}\epsilon_{\mu\nu\rho \sigma}P^{\sigma \rho}$. By using the Legendre transformation of the Lagrangian, the corresponding functional structure \( H(P, O) \) is derived as
\begin{align}
H(P, O) = -\frac{1}{2} P_{\mu\nu} F^{\mu\nu} - \mathcal{L}.
\end{align}
In the $\mathit{EH}$ model, the functional simplifies to
\begin{align}
H(P, O) = P - \frac{a}{2} P^2 - \frac{7a}{8} O^2.
\end{align}
The field equations for the \( F(R) \)-$\mathit{EH}$ theory of gravity are obtained by varying the action with respect to the metric tensor and field strength. These yield the following equations
\begin{align} \label{equa}
8\pi T_{\mu\nu} = R_{\mu\nu}(1 + f_R) - \frac{g_{\mu\nu} F(R)}{2} + (g_{\mu\nu} \nabla^2 - \nabla_\mu \nabla_\nu) f_R.
\end{align}
The field strength equation (for the \( P \)-setting) given as
\begin{align}\label{equb}
\nabla_\mu P^{\mu\nu} = 0,
\end{align}
where \( f_R = \frac{d f(R)}{dR} \) and \( T_{\mu\nu} \) is the energy-momentum tensor related to the $\mathit{EH}$ term in the \( P \)-frame follow as 
\begin{align}
T_{\mu\nu} = \frac{1}{4\pi} \left( (1 - a P) P^{\beta}_\mu P_{\nu\beta} + g_{\mu\nu} \left( P - \frac{3}{2} a P^2 - \frac{7a}{8} O^2 \right) \right).
\end{align}
The next step is to solve the field equations (\ref{equa}) and (\ref{equb}) under the assumption of a static, spherically symmetric spacetime metric defined by
\end{justify}
\begin{align} \label{metrica}
    ds^2= -A(r)dt^2+\frac{dr^2}{A(r)}+r^2(d\theta^2+\sin^2\theta d\phi^2).
\end{align}

From a practical standpoint, solving the field equations in the framework of
F(R) gravity coupled with $\mathit{EH}$ matter fields, as given in Eq.~\eqref{equa}, poses significant challenges. Finding exact analytical solutions under these conditions is notably difficult. To address this complication, one can impose the condition that the energy-momentum tensor of the EH field is traceless. Under this assumption, it becomes possible to derive analytical solutions within 
F(R) gravity when coupled to an NLED field. Therefore, obtaining a black hole solution with constant curvature in this modified gravity setting requires that the trace of the stress-energy tensor $ T_{\mu\nu} $ vanishes \cite{Moon2011fr}. Assuming a constant scalar curvature $ R=R_{0} $, as suggested in \cite{sekhmani2024electrically}, the trace of Eq.~\eqref{equa} simplifies as
\begin{equation}
R_{0}\left( 1+f_{R_0}\right) -2\left( R_{0}+f(R_0) \right) =0,
\label{R0}
\end{equation}
where $ f_{R_0}=f_{R_{\vert R=R_{0}}} $. Consequently, the constant scalar curvature can be determined from Eq.~\eqref{R0} such that:
\begin{equation}
R_{0}=\frac{2f(R_0)}{f_{R_0}-1}.
\label{R01}
\end{equation}
By substituting Eq.\eqref{R01} into Eq.\eqref{equa}, the field equations of the $F(R)$ -EH theory can be properly formulated as
\begin{equation}
    R_{\mu\nu}(1+f_{R_0})-\frac{g_{\mu\nu} }{4}R_0(1+f_{R_0})=8\pi T_{\mu\nu}.\label{moe}
\end{equation}
Specifically, the parameter $ f_{R_0} $ allows for a smooth reduction to GR in the limit 
$f_{R_0}=0$. In this case, the field equation \eqref{equa} reduces to the form consistent with the GR-$ \Lambda $-EH
Theory of gravity. 
Regarding the electromagnetic field, only the electric component of the EH theory is considered, allowing the electromagnetic field tensor to be explicitly derived as
\begin{equation}
    P_{\mu\nu}=\frac{q}{r^2}\delta^0_{[\mu}\delta^1_{\nu]}. \label{pp}
\end{equation}
In which the electromagnetic invariants are defined as
\begin{equation}
    P=\frac{q^2}{2r^4},\quad O=0,
\end{equation}
where $q$ denotes an integration constant related to the electric charge. By using the metric function \eqref{metrica}, the electric field tensor \eqref{pp}, and solving the field equations \eqref{equa}, a specific solution for the constant scalar curvature 
$(R = R_0= \text{constant})$ can be obtained as
\begin{align}\label{solmetric}
A(r) = 1 - \frac{m_0}{r} - \frac{R_0r^2} {12}+ \frac{1  }{1 + f_{R_0}}  \left( \frac{q^2}{r^2} - \frac{a q^4}{20r^6} \right),
\end{align}
where \( m_0 \) is an integration constant that represents the black hole’s geometric mass. Moreover, the charge parameter \( q \) can be interpreted physically through a closed surface integral concept using the conservation Gauss's law \cite{sekhmani2024electrically}. These insights into the resulting black hole solution (\ref{solmetric}) suggest that it effectively describes the field equations in the context of the \( F(R) \)-$\mathit{EH}$ theory of gravity. 
To better handle the black hole solution, certain criteria must be satisfied, such as ensuring that \( f_{R_0} \neq -1 \). It is worth noting that by restricting to a specific subspace of the parameter space, namely $f_{R_0}=0$, $R_0=4\Lambda$, and $a=0$, one recovers the well-known Reissner–Nordstrom–(A)dS black hole, defined by
\begin{equation}
\label{solmetricrn}
   A(r)=1-\frac{m_{0}}{r}-\frac{\Lambda r^2}{3} +\frac{q^2}{r^2}-\frac{a q^4}{20 r^6}.
\end{equation}

A comparison between the metric functions \eqref{solmetric} and \eqref{solmetricrn} reveals that the spacetime becomes asymptotically (A)dS when $R_0=4\Lambda$, and $ \Lambda >0 $ ($ \Lambda <0 $). 
\section{THE EFFECTIVE GEOMETRY FOR the F(R)-{EH} BLACK HOLE}\label{sec3}
In linear electrodynamics, electromagnetic waves do not interact with electrostatic fields. However, in a non-linear framework, photons are influenced by the non-linearity of the field and follow null geodesics defined by an effective geometry, rather than those of Minkowski spacetime. This implies that in a complex vacuum, light behaves as if it propagates through a medium that modifies its trajectory.
To describe the effective geometry arising from F(R)-$\mathit{EH}$-NLED, we adopt the method given in \cite{kruglov2020shadow}. Since the Lagrangian considered is a single-parameter function, the corresponding equation of motion is given by
\begin{align}\label{equca}
    \partial_\mu (\mathcal{L}_F F^{\mu \nu})=0.
\end{align}
The null geodesics of photon paths are governed by an effective metric defined as
\begin{equation}\label{equcb}
g_{\text{eff}}^{\mu \nu} = \mathcal{L}_F g^{\mu \nu}- 4 \mathcal{L}_{FF} F_\alpha ^{\mu} F ^{\alpha \nu},
\end{equation} 
where, $\mathcal{L}_F = \frac{d\mathcal{L}}{dF}, \quad \mathcal{L}_{FF} = \frac{d^2\mathcal{L}}{dF^2}$.
Using (\ref{equca}) and (\ref{equcb}) along with the metric element from (\ref{metrica}), the effective metric for the electrically charged F(R)-$\mathit{EH}$ black hole takes the form
\begin{equation}
ds^2_{\text{eff}} = K(r) [A(r) dt^2 + A^{-1}(r) dr^2 ]+ k(r) r^2 (d\theta^2 + \sin^2\theta \, d\phi^2).
\end{equation}
The functions \( K(r) \) and \(k(r)\) are given by
\begin{equation}
K(r) = 1 + \frac{a(aq^3 - 2qr^4)^2}{8r^{12}}, \quad
{k}(r) = 1 + \frac{a(aq^3 - 2qr^4)^2}{8r^{12}} - \frac{aq^2}{r^4}.
\end{equation}
The functions 
$ K(r) $ and $ k(r) $ must remain positive to ensure that the effective geometry preserves its signature during photon motion. This effective metric is the foundation for studying the shadow structure of F(R)-$\mathit{EH}$ black holes, which we will address in the following section.
\section{Optical properties}\label{sec4}
In this section, we explore the optical characteristics of electrically charged black holes in \( F(R) \)- $\mathit{EH}$ theory. These include the trajectory of light rays, shadow formation, energy emission, and observational constraints from EHT. The optical properties, particularly photon spheres and shadows, are shaped by the interplay between the modifications from \( F(R) \) gravity and the nonlinear effects of $\mathit{EH}$ electrodynamics. By modeling these black holes as supermassive objects, we also aim to place constraints on the free parameters of the theory. This is done by comparing with astrophysical observations, especially the shadow profile captured for the black hole at the center of M87$\ast$.
\subsection{Null geodesic and Light bending}\label{sec4a}
We now turn our attention to the geodesic structure of photons in a spherically symmetric spacetime. In the context of NLED, photons do not follow the null geodesics of the background metric but instead move along null geodesics defined by the effective geometry. The Lagrangian describing photon motion in the spacetime of an electrically charged black hole in \( F(R) \)-$\mathit{EH}$ theory is given by:

\begin{equation}
\mathfrak{L}(\dot{x}^a) = \frac{1}{2} g_{\mu\nu}^{\text{eff}} \dot{x}^\mu \dot{x}^\nu = \frac{1}{2} \left[ A(r) K(r) \dot{t}^2 - A(r) K^{-1}(r) \dot{r}^2 - k(r) r^2 (\dot{\theta}^2 + \sin^2\theta \, \dot{\phi}^2) \right],
\end{equation}
where the overdot denotes differentiation with respect to an affine parameter along the geodesic. Due to the spherical symmetry of the black hole, photon trajectories can be confined to the equatorial plane \( \theta = \frac{\pi}{2} \), simplifying the Lagrangian to
\begin{equation}
\mathfrak{L} = \frac{1}{2} \left[ A(r) K(r) \dot{t}^2 - \frac{A(r)}{K(r)} \dot{r}^2 - k(r) r^2 \dot{\phi}^2 \right],
\end{equation}
From the Lagrangian, the canonical momenta corresponding to the coordinates \( t \), \( r \), and \( \phi \) are derived as
\begin{align}
p_t &= \frac{\partial \mathfrak{L}}{\partial \dot{t}} = K(r) A(r) \dot{t} = E,  \\
p_r &= \frac{\partial \mathfrak{L}}{\partial \dot{r}} = \frac{K(r)}{A(r)} \dot{r},  \\
p_\phi &= \frac{\partial \mathfrak{L}}{\partial \dot{\phi}} = k(r) r^2 \dot{\phi} = L .
\end{align}
Here, \( E \) and \( L \) denote the conserved energy and angular momentum of the photon, respectively. Using the Lagrangian formalism, one can derive three first-order differential equations that describe the geodesic motion
\begin{align}
\frac{dt}{d\lambda} &= \frac{1}{b K(r)} \cdot \frac{1}{A(r)},  \\
\frac{dr}{d\lambda} &= \sqrt{\frac{1}{b^2 K(r)^2} - \frac{A(r)}{K(r) k(r) r^2}}, \\
\frac{d\phi}{d\lambda} &= \frac{1}{k(r) r^2} .
\end{align}
where \( \lambda \) is the affine parameter and \( b = \frac{L}{E} \) is the impact parameter, which represents the closest approach of a light ray to the black hole.
From the Lagrangian, the equation of motion for a photon moving along a null geodesic is given by
\begin{equation}\label{equcc}
K(r) A(r) \dot{t}^2 - \frac{K(r)}{A(r)} \dot{r}^2 - r^2 k(r) \dot{\phi}^2 = 0.
\end{equation}
Using (\ref{equcc}), the effective potential \( \mathrm{V}_{eff}\) governing photon motion can be expressed as
\begin{equation}\label{equcd}
\left( \frac{dr}{d \phi} \right)^2 = V_{eff} = r^4\left(\frac{ k(r)^2}{b^2 K(r)^2} - \frac{A(r) k(r)}{K(r) r^2}\right).
\end{equation}
The behavior of photon trajectories near a black hole is determined by their initial conditions, especially the value of the impact parameter. In regions of strong gravity, photons can follow circular orbits at specific radii, this defines the photon sphere. The radius of the photon sphere, \( r_{\text{p}} \), can be derived using the effective potential. If a photon’s path slightly deviates from the photon sphere, it will either be captured by the black hole or escape to infinity. The condition for an unstable circular orbit is given by the simultaneous requirements \(
\frac{dV_{eff}}{dr} = 0 \quad \text{and} \quad V_{eff} = 0.
\) Applying this to (\ref{equcd}) leads to
\begin{equation}\label{equce}
\frac{1}{b^2} = \frac{E^2}{L^2} = \frac{A(r) K(r)}{r^2 k(r)}.
\end{equation}
To determine the radius of the photon sphere \( r_{\text{p}} \), we apply the condition for an unstable circular orbit. This leads to the following equation
\begin{equation}\label{equcf}
r A'(r) K(r) k(r) + 2 A(r) K(r) k(r) - r A(r) k'(r) K(r) - r K'(r) k(r) A(r) = 0.
\end{equation}
Substituting the photon sphere radius \( r_{\text{p}} \) obtained from (\ref{equcf}) into (\ref{equce}), we determine the critical impact parameter \( b_c \) as
\begin{equation}
b_c^2 = \frac{K(r_{\text{p}}) A(r_{\text{p}})}{k(r_{\text{p}}) r_{\text{p}}^2}.
\end{equation}
Introducing a new variable \( u = \frac{1}{r} \), the orbit equation becomes
\begin{equation}
\left( \frac{du}{d\phi} \right)^2 = H(u) = \frac{k(u)^2}{b^2 K(u)^2} - \frac{u^2 A(u) k(u)}{K(u)}.
\end{equation}
The function \( H(u) \) determines the photon’s motion depending on the value of the impact parameter \( b \). Photon trajectories near the black hole fall into three categories are given by
\begin{itemize}
\item If \( b > b_c \), the photon approaches the black hole, reaches a minimum distance, and then returns to infinity. The turning point corresponds to the smallest positive real root of \( H(u) = 0 \), denoted \( u_m \). The total deflection angle is

\begin{equation}
\Delta\phi = 2 \int_0^{u_m} \frac{du}{\sqrt{H(u)}}.
\end{equation}
\item If \( b < b_c \), the photon falls into the black hole without returning. The corresponding deflection angle is
\begin{equation}
\Delta\phi = \int_0^{u_0} \frac{du}{\sqrt{H(u)}}, \quad \text{where} \quad u_0 = \frac{1}{r_0}.
\end{equation}
\item If \( b = b_c \), the photon orbits the black hole at the photon sphere radius.
\end{itemize}
According to \cite{yu1995some}, photon trajectories can be classified into three types: direct, lensed, and photon ring paths, depending on their behavior near the black hole. The total number of orbits completed before the photon escapes or is absorbed can be characterized by
\begin{equation}\label{equcga}
n(b) = \frac{2m - 1}{4}, \quad m = 1, 2, 3, \ldots
\end{equation}
For each \( m \), (\ref{equcga}) yields two solutions for the impact parameter: \( b_{m}^{-}\) and \( b^+_m \), representing the minimum and maximum values, respectively. These satisfy the inequality \( b_{m}^{-} < b_c < b^+_m \), helping classify all photon trajectories based on their orbital characteristics.
\begin{table}[ht]
    \centering
    \scriptsize
    \caption{Regions of direct rays, lensing rings, and photon rings with $a = 0.1$ and different values of $q$, $R_0$, and $f_{R_0}$.}
    \label{table1}
    \begin{tabular}{|c|c|c|c|c|}
        \hline\hline
        Parameters & 
        \makecell{$q = 0.2$, $f_{R_0} = 0.1$ \\ $R_0 = -0.02$} & 
        \makecell{$q = 0.5$, $f_{R_0} = 0.1$ \\ $R_0 = -0.02$} & 
        \makecell{$q = 0.2$, $f_{R_0} = 0.1$ \\ $R_0 = -0.08$} & 
        \makecell{$q = 0.2$, $f_{R_0} = -0.5$ \\ $R_0 = -0.02$} \\
        \hline
        Direct rays 
        & $b < 2.426$ 
        & $b < 1.925$ 
        & $b < 2.392$ 
        & $b < 2.374$ \\
        $\left(n < \frac{3}{4}\right)$ 
        & $b > 3.006$ 
        & $b > 2.677$ 
        & $b > 2.942$ 
        & $b > 2.966$ \\
        \hline
        Lensing rings 
        & $2.426 < b < 2.515$ 
        & $1.925 < b < 2.065$ 
        & $2.392 < b < 2.476$ 
        & $2.374 < b < 2.469$ \\
        $\left(\frac{3}{4} < n < \frac{5}{4}\right)$ 
        & $2.536 < b < 3.006$ 
        & $2.109 < b < 2.677$ 
        & $2.496 < b < 2.942$ 
        & $2.492 < b < 2.966$ \\
        \hline
        Photon ring $\left(n > \frac{5}{4}\right)$ 
        & $2.515 < b < 2.536$ 
        & $2.065 < b < 2.109$ 
        & $2.476 < b < 2.496$ 
        & $2.469 < b < 2.492$ \\
        \hline
    \end{tabular}
\end{table}
The classification of photon trajectories based on the number of orbits \( n \) and the impact parameter \( b \) is as follows

\begin{itemize}
\item \textbf{Direct:} For \( n < \frac{3}{4} \), the impact parameter lies in the range \( b \in (b_1^{-}, b_2^{-}) \cup (b_2^+,\infty) \). In this case, the photon trajectory intersects the equatorial plane only once.

\item \textbf{Lensed:} For \( \frac{3}{4} < n < \frac{5}{4} \), the impact parameter lies in the range \( b \in (b_2^{-}, b_3^{-}) \cup (b_3^+, b_2^+) \). Here, the photon intersects the equatorial plane at least twice, resulting in a lensing effect.

\item \textbf{Photon ring:} For \( n > \frac{5}{4} \), the impact parameter satisfies \( b \in (b_3^{-}, b_3^+) \). In this regime, the photon trajectory crosses the equatorial plane at least three times, leading to the formation of a photon ring.
\end{itemize}
\begin{figure}[!htb]
    \centering
    \subfloat[$q = 0.2$, $R_0 = -0.02$, $f_{R_0} = 0.1$]{
        \includegraphics[width=0.33\textwidth]{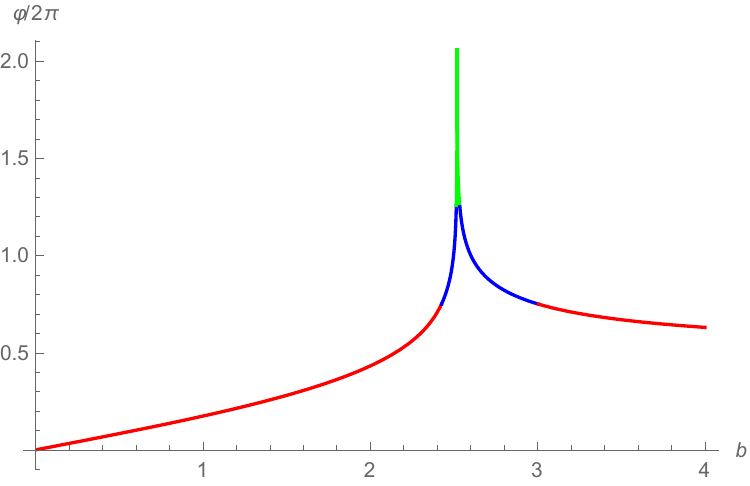}
    }
    \subfloat[$q = 0.2$, $R_0 = -0.02$, $f_{R_0} = 0.1$]{
        \includegraphics[width=0.26\textwidth]{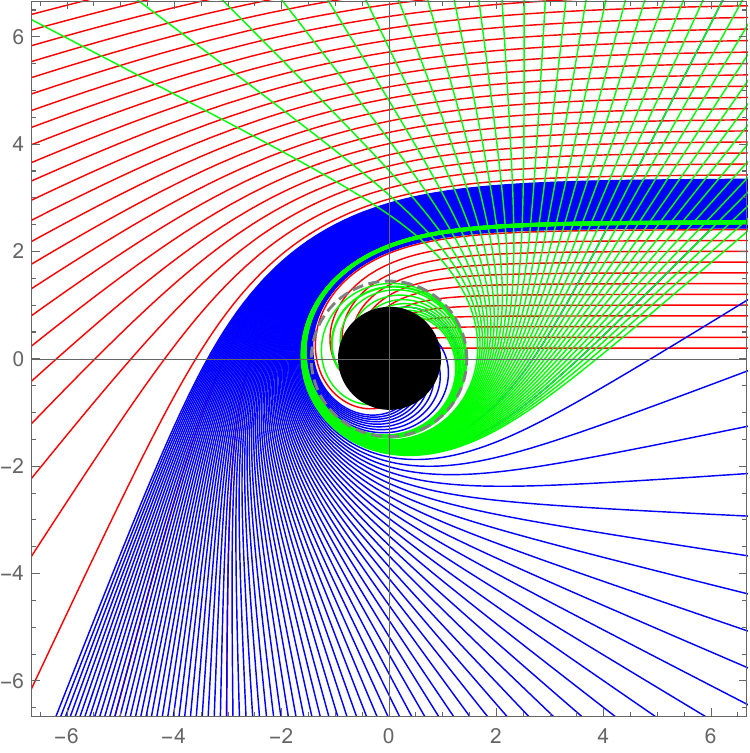}
    }\\[1ex]
    
    \subfloat[$q = 0.5$, $R_0 = -0.02$, $f_{R_0} = 0.1$]{
        \includegraphics[width=0.33\textwidth]{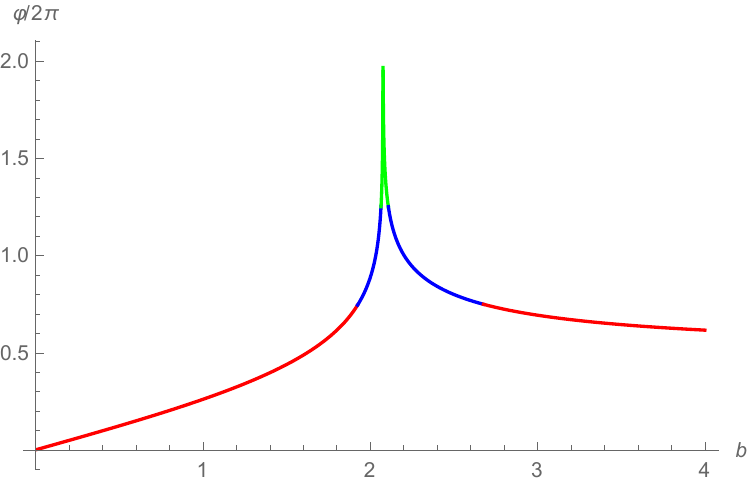}
    }
    \subfloat[$q = 0.5$, $R_0 = -0.02$, $f_{R_0} = 0.1$]{
        \includegraphics[width=0.26\textwidth]{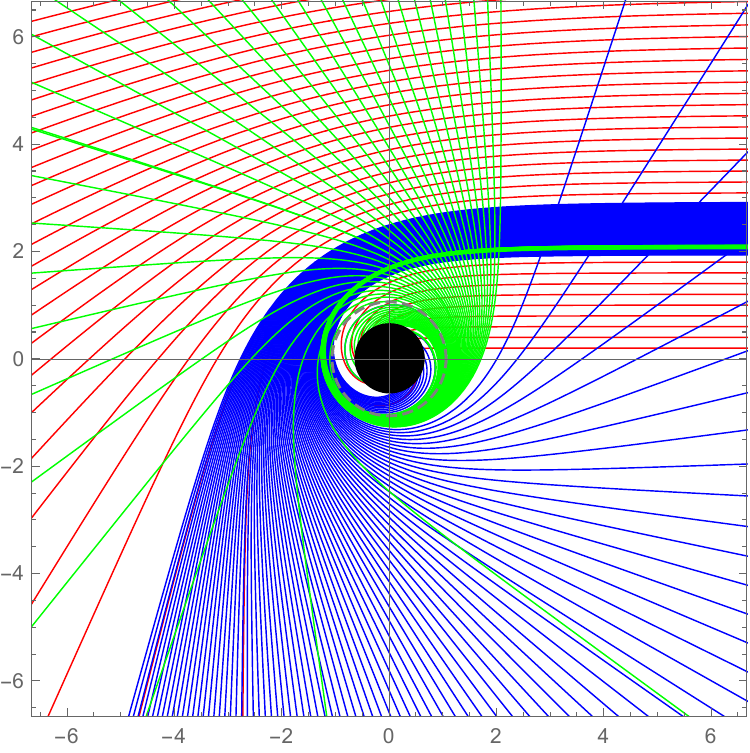}
    }\\[1ex]
    
    \subfloat[$q = 0.2$, $R_0 = -0.08$, $f_{R_0} = 0.1$]{
        \includegraphics[width=0.33\textwidth]{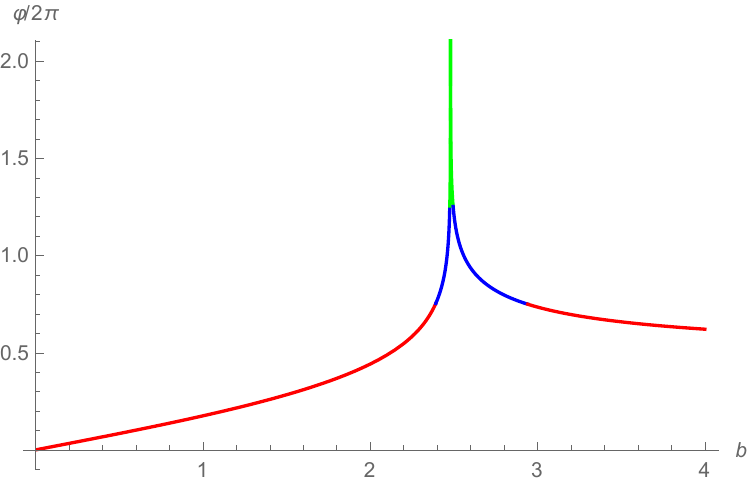}
    }
    \subfloat[$q = 0.2$, $R_0 = -0.08$, $f_{R_0} = 0.1$]{
        \includegraphics[width=0.26\textwidth]{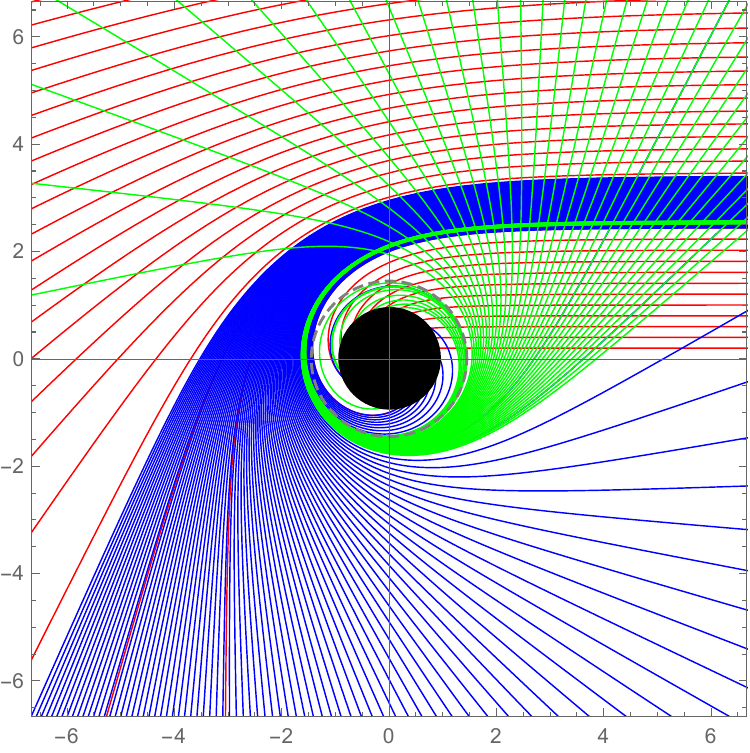}
    }\\[1ex]
    
    \subfloat[$q = 0.2$, $R_0 = -0.02$, $f_{R_0} = -0.5$]{
        \includegraphics[width=0.33\textwidth]{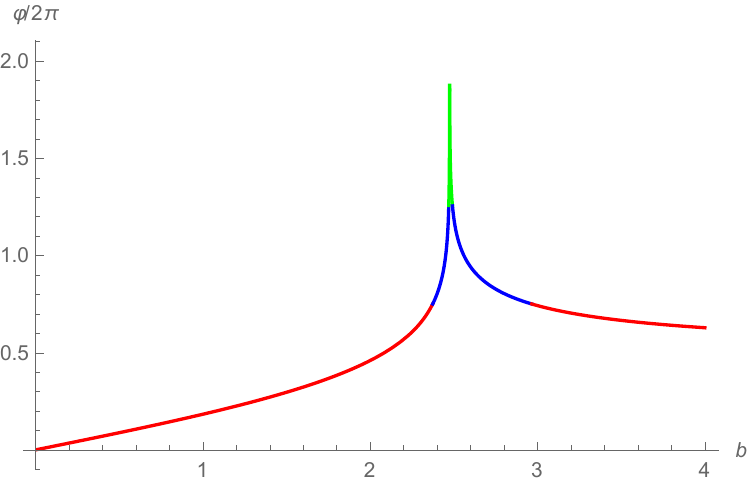}
    }
    \subfloat[$q = 0.2$, $R_0 = -0.02$, $f_{R_0} = -0.5$]{
        \includegraphics[width=0.26\textwidth]{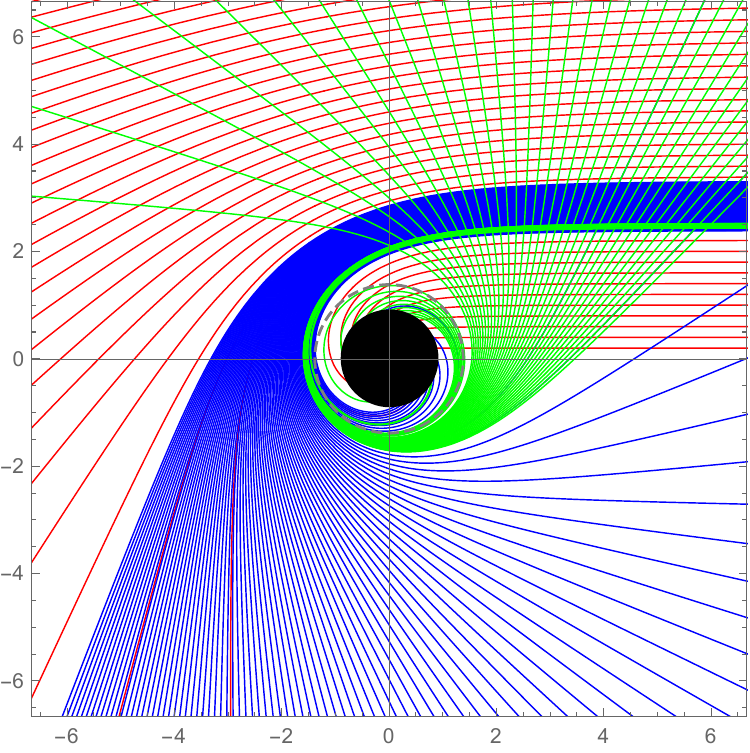}
    }

    \caption{The number of photon orbits $n$ (left column) and trajectories of photons (right column) as a function of the impact parameter $b$ for the F(R)-EH-AdS black hole with $a = 0.1$. The red lines, blue lines, and green lines correspond to $n < \frac{3}{4}$, $\frac{3}{4} < n < \frac{5}{4}$, and $n > \frac{5}{4}$, respectively. The black disk and the dashed curves denote the event horizon and photon sphere.}
    \label{Figphoton}
\end{figure}
\begin{figure}[!htb]
    \centering

    % Top row: de Sitter
    \subfloat[]{
        \includegraphics[width=0.31\textwidth]{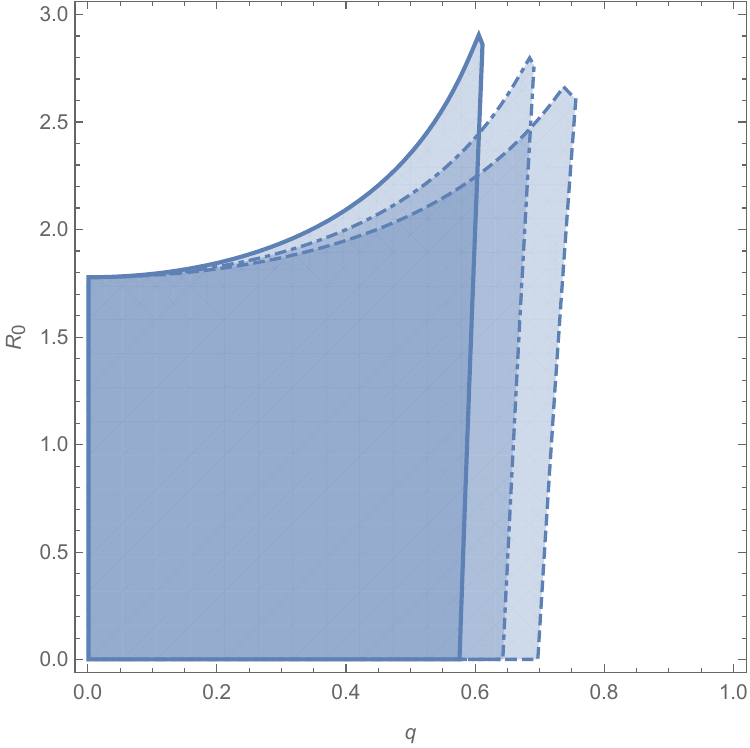}
    }\hfill
    \subfloat[]{
        \includegraphics[width=0.31\textwidth]{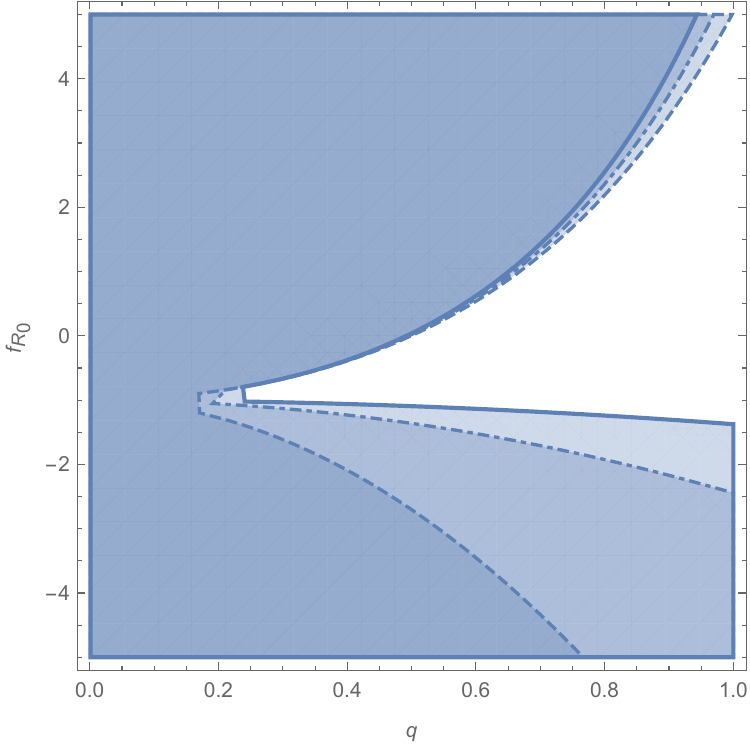}
    }\hfill
    \subfloat[]{
        \includegraphics[width=0.31\textwidth]{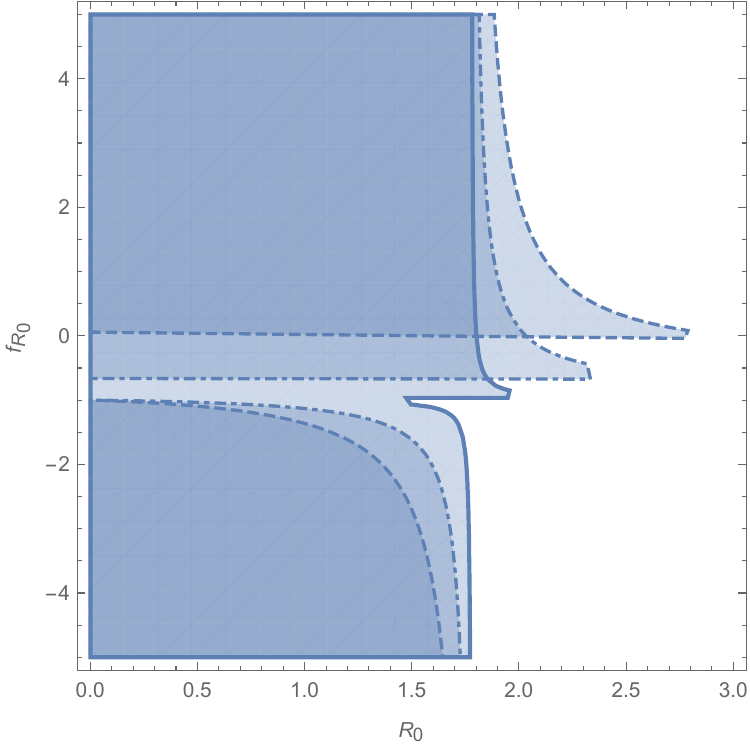}
    }\\[1.5ex]

    % Bottom row: AdS
    \subfloat[]{
        \includegraphics[width=0.31\textwidth]{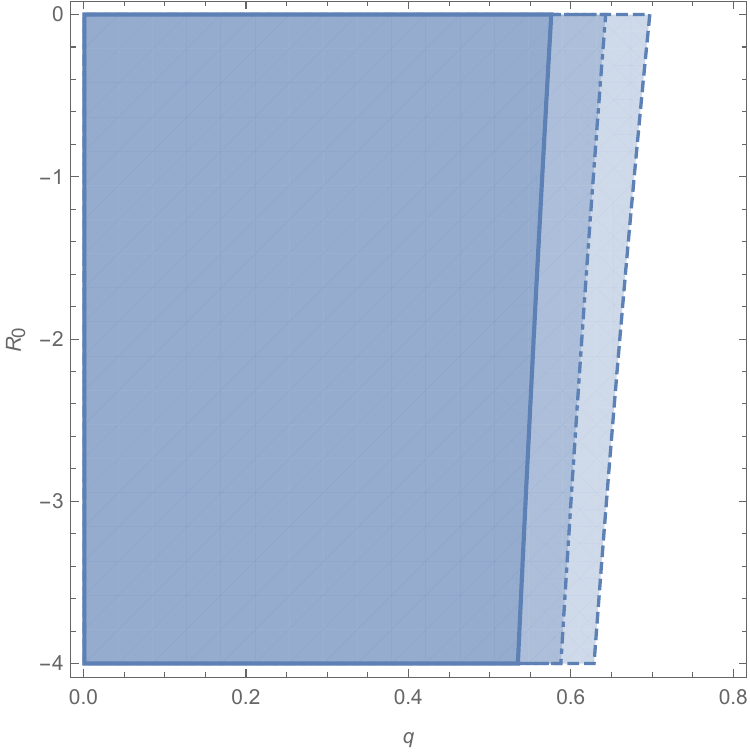}
    }\hfill
    \subfloat[]{
        \includegraphics[width=0.31\textwidth]{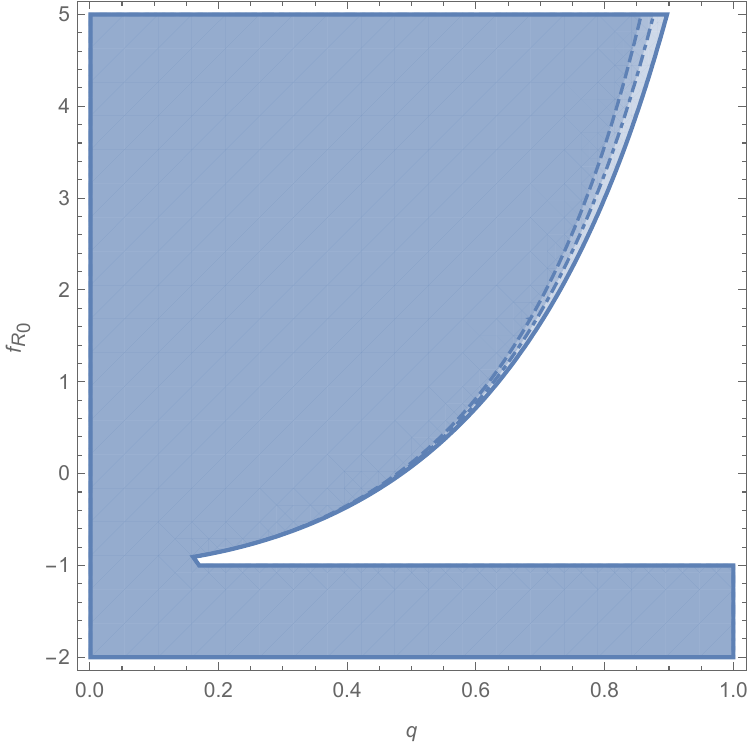}
    }\hfill
    \subfloat[]{
        \includegraphics[width=0.31\textwidth]{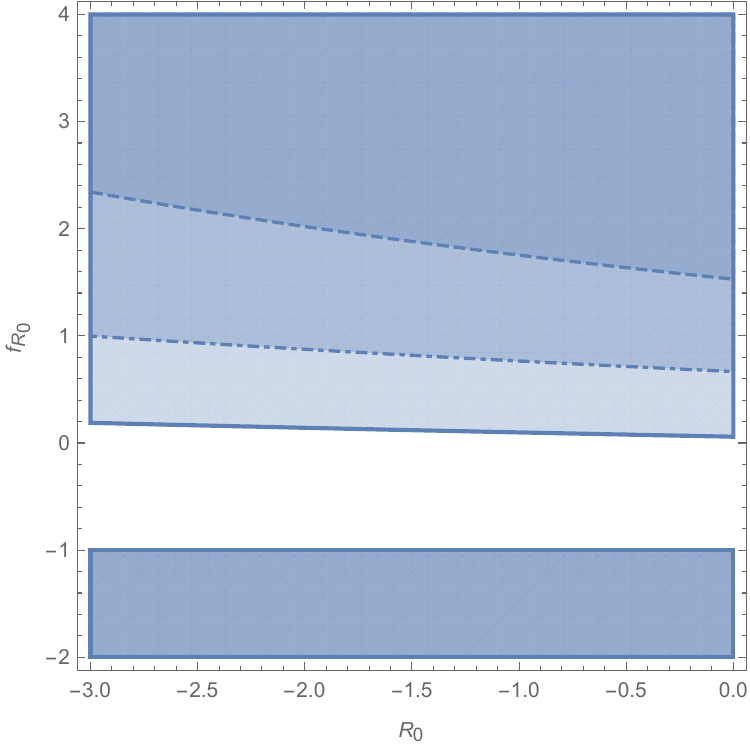}
    }
    \caption{\textbf{Top row:} The constraint $\frac{r_{p}}{r_{eh}} > 1$ (shaded areas) for the F(R)-EH-dS black hole is shown in:
    (a) $(R_{0},q)$ plane for $f_{R_{0}} = 0.5$ (solid), $f_{R_{0}} = 1.0$ (dash-dotted), and $f_{R_{0}} = 1.5$ (dashed); 
    (b) $(f_{R_{0}},q)$ plane for $R_{0} = 0.5$, $1.0$, and $1.5$; 
    (c) $(f_{R_{0}},R_{0})$ plane for $q = 0.1$, $0.3$, and $0.5$.
    \textbf{Bottom row:} The constraint $\frac{r_{p}}{r_{eh}} > 1$ for the F(R)-EH-AdS black hole is shown in:
    (d) $(R_{0},q)$ plane for $f_{R_{0}} = 0.5$, $1.0$, and $1.5$; 
    (e) $(f_{R_{0}},q)$ plane for $R_{0} = -0.5$, $-1.0$, and $-1.5$; 
    (f) $(f_{R_{0}},R_{0})$ plane for $q = 0.5$, $0.6$, and $0.7$. The unshaded region corresponds to $\frac{r_{p}}{r_{eh}} < 1$, which is physically forbidden. Parameters are set as $M = 1$, $a = 0.2$.}
    \label{Fig1}
\end{figure}
\begin{figure}[!htb]
    \centering

    % Top row: dS
    \subfloat[]{
        \includegraphics[width=0.31\textwidth]{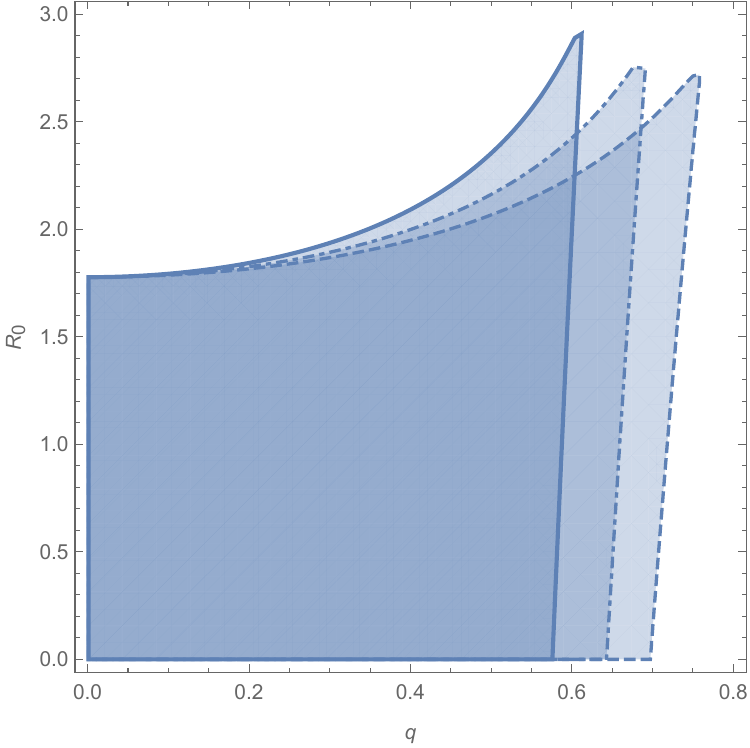}
    }\hfill
    \subfloat[]{
        \includegraphics[width=0.31\textwidth]{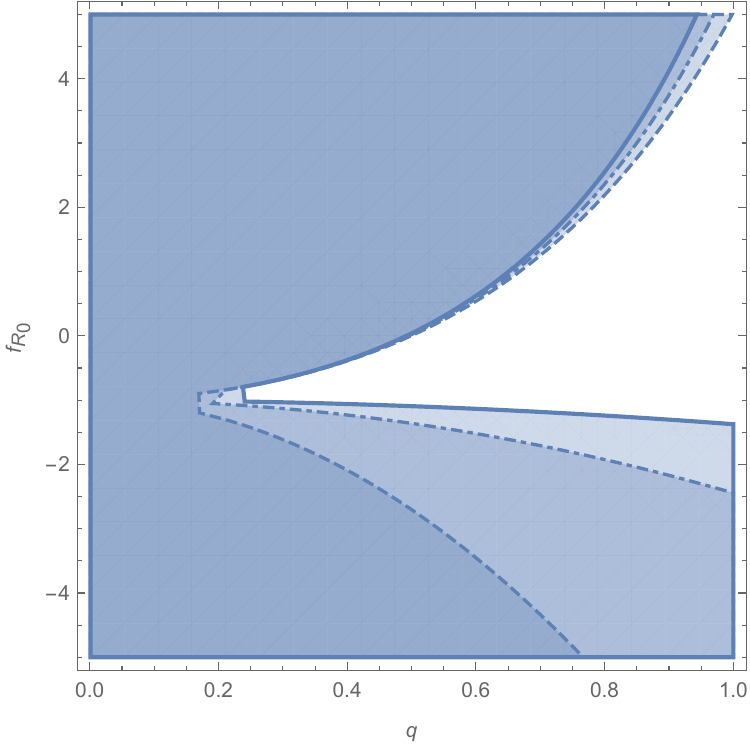}
    }\hfill
    \subfloat[]{
        \includegraphics[width=0.31\textwidth]{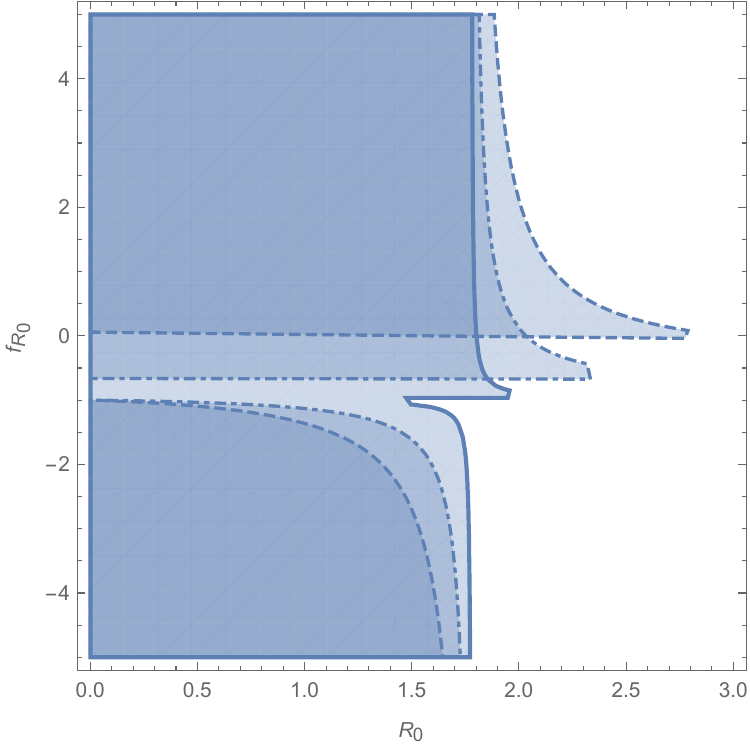}
    }\\[1.5ex]

    % Bottom row: AdS
    \subfloat[]{
        \includegraphics[width=0.31\textwidth]{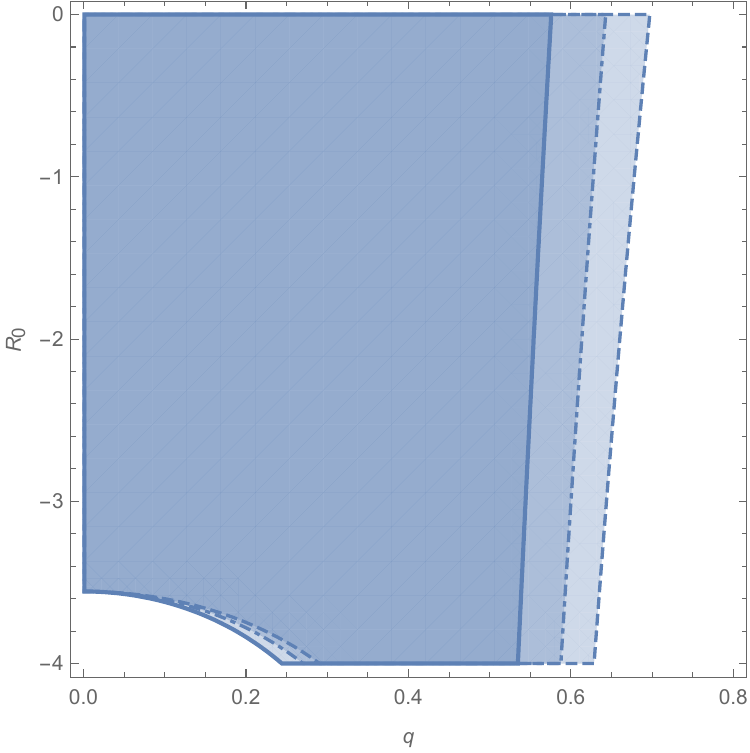}
    }\hfill
    \subfloat[]{
        \includegraphics[width=0.31\textwidth]{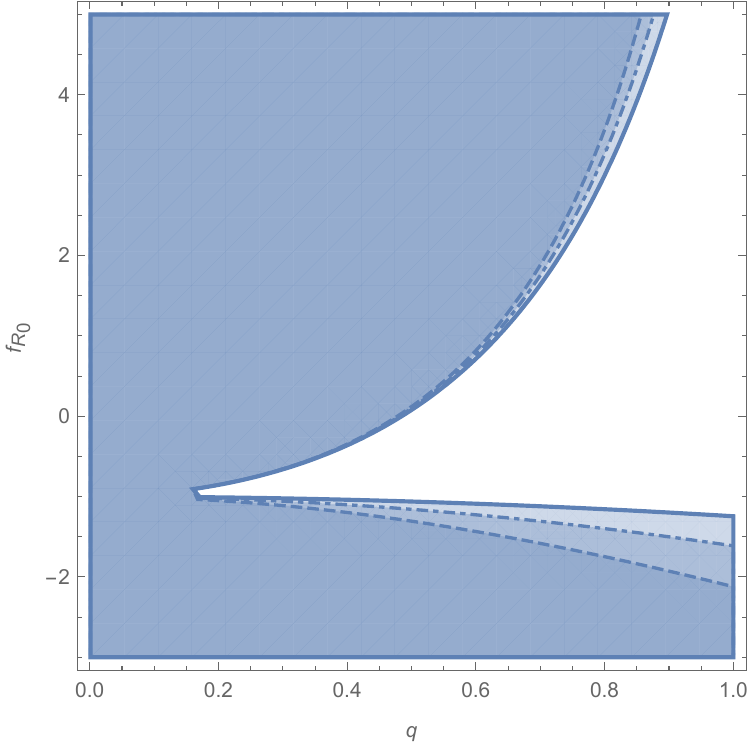}
    }\hfill
    \subfloat[]{
        \includegraphics[width=0.31\textwidth]{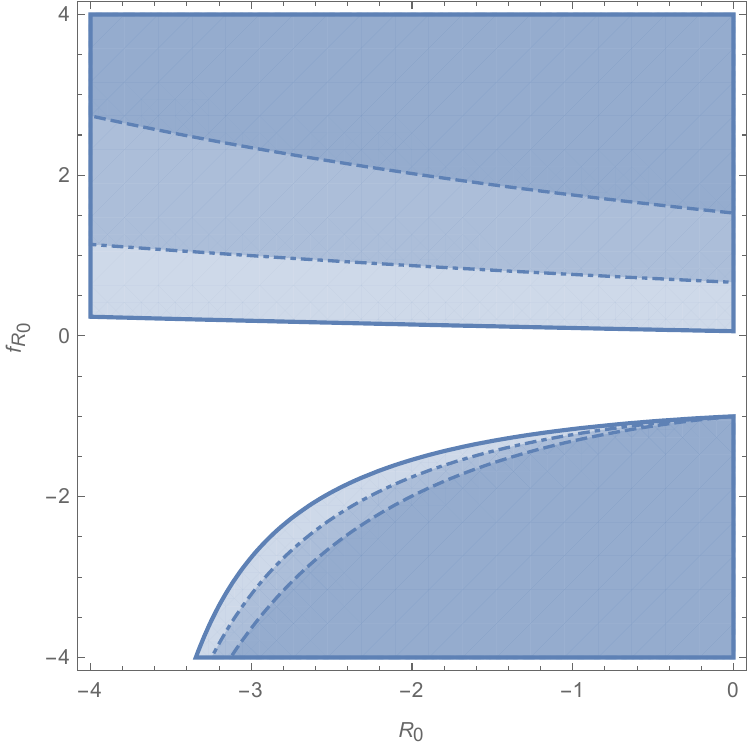}
    }
    \caption{\textbf{Top row:} The constraint $\frac{r_{sh}}{r_{p}} > 1$ (shaded areas) for the F(R)-EH-dS black hole is shown in:
    (a) $(R_0, q)$ plane for $f_{R_0} = 0.5$ (solid), $f_{R_0} = 1.0$ (dash-dotted), and $f_{R_0} = 1.5$ (dashed); 
    (b) $(f_{R_0}, q)$ plane for $R_0 = 0.5$, $1.0$, and $1.5$; 
    (c) $(f_{R_0}, R_0)$ plane for $q = 0.1$, $0.3$, and $0.5$.
    \textbf{Bottom row:} The constraint $\frac{r_{sh}}{r_{p}} > 1$ for the F(R)-EH-AdS black hole is shown in:
    (d) $(R_0, q)$ plane for $f_{R_0} = 0.5$, $1.0$, and $1.5$; 
    (e) $(f_{R_0}, q)$ plane for $R_0 = -0.5$, $-1.0$, and $-1.5$; 
    (f) $(f_{R_0}, R_0)$ plane for $q = 0.5$, $0.6$, and $0.7$. The unshaded (colorless) regions correspond to $\frac{r_{sh}}{r_{p}} < 1$, which violates the condition. Parameters: $M = 1$, $a = 0.2$.}
    \label{Fig2}
\end{figure}
Fig.~\ref{Figphoton} shows the number of photon orbits \( n \) (left column) and the corresponding photon trajectories (right column) as functions of the impact parameter \( b \). In this figure, red lines represent direct emission rays, blue indicate lensed rays, and green correspond to photon ring rays. The photon orbit is marked by a dashed gray circle, and the black disk indicates the black hole event horizon. From the left panels, it is evident that as the impact parameter \( b \) approaches the critical value \( b_c \), the photon orbit number \( n \) exhibits a sharp peak. Beyond this critical threshold, all photon paths correspond to direct emission.

A comparison between the different panels in Fig.~\ref{Figphoton} reveals that both the electric charge \( q \) and the parameter \( f_{R_0} \), which arises from the \( F(R) \)-$\mathit{EH}$ model, significantly influence the classification of photon paths. As supported by the data in Table~\ref{table1} and the visual trends in Fig.~\ref{Figphoton}, increasing the charge \( q \) and the absolute value of \( f_{R_0} \)  result in broader ranges for lensed and photon ring trajectories, while increasing \( R_0 \) leads to decreasing the range of lensing and photon rings. These parameters directly shape the effective geometry defined by the metric function (\ref{solmetric}), which governs the behavior of null geodesics and, consequently, the black hole shadow and photon dynamics.
\subsection{Shadow of the F(R)-{EH}  black hole}\label{sec4b}
The black hole shadow is the apparent silhouette formed in the sky as seen by a distant observer, resulting from the bending and capture of light by the strong gravitational field near the event horizon. It appears as a dark, two-dimensional region superimposed on a bright background, typically from an accretion disk or surrounding light sources. The boundary of the shadow delineates the photon region, distinguishing light rays that are absorbed by the black hole from those that escape to infinity. In spherically symmetric spacetimes, this photon region reduces to a well-defined surface known as the photon sphere. In the context of electrically charged black holes in \( F(R) \)-$\mathit{EH}$ theory, the shadow radius can be determined using the effective potential defined in (\ref{equcd}) and the photon sphere radius obtained from (\ref{equcf}). The shadow radius is then given by
\begin{align}\label{equcg}
R_{\text{sh}} = r_{\text{p}}\sqrt{ \frac{k(r_{\text{p}})}{ K(r_{\text{p}}) A(r_{\text{p}}) }}.
\end{align}
To ensure a physically viable optical structure, the condition \(
r_{\text{eh}} < r_{\text{p}} < R_{\text{sh}}
\) must be satisfied, where \( r_{\text{eh}} \) is the radius of the event horizon. This condition guarantees that the shadow is well-defined and observable. Analyzing such constraints allows us to identify the admissible parameter space in which the model produces realistic shadow features. 
 Fig.~\ref{Fig1} displays the admissible parameter space that satisfies the condition \( r_{\text{p}} / r_{\text{eh}} > 1 \), ensuring that the photon sphere lies outside the event horizon. The shaded regions correspond to parameter combinations that yield physically viable black hole solutions. In the top row, Figs.~\ref{Fig1}(a)–\ref{Fig1}(c) present the admissible regions in the \( (R_0, q) \), \( (f_{R_0}, q) \), and \( (f_{R_0}, R_0) \) planes, respectively, for fixed values of \( f_{R_0} \), \( R_0 \), and \( q \) in dS spacetime (positive \( R_0 \)).
As seen in Fig.~\ref{Fig1}(a), increasing the electric charge \( q \) enlarges the admissible region, indicating that a stronger electric field enhances the conditions for forming a photon sphere outside the horizon.  
Similarly, from Fig.~\ref{Fig1}(b), we observe that increasing \( \vert f_{R_0} \vert \) expands the admissible parameter space, reflecting the influence of $ F(R) $ gravity. As for the effect of \( q \) and \( f_{R_0} \) on the admissible parameter space of AdS black holes (negative \( R_0 \)), the bottom panels of Fig.~(\ref{Fig1}) verify that their effect is similar to what is being stated for dS ones. The only difference is that the influence of the parameter \( f_{R_0} \) on the allowed regions is significant in dS spacetime (compare Fig.~\ref{Fig1}(b) and Fig.~\ref{Fig1}(e)), while the electric charge has a substantial effect in AdS spacetime (compare Fig.~\ref{Fig1}(c) and Fig.~\ref{Fig1}(f)). 
To examine the condition \( R_{\text{sh}} / r_{\text{p}} > 1 \), which ensures that the shadow forms outside the photon sphere, Fig.~\ref{Fig2} presents the corresponding admissible regions. Similar to the previous trend, Fig.~\ref{Fig2}(a) shows that the electric charge \( q \) has a positive effect, broadening the range of valid parameters.  Fig.~\ref{Fig2}(b) confirms that a larger parameter \(  f_{R_0} \) enhances the allowed region, increasing the configurations that lead to physically meaningful black hole shadows. Comparing Fig.~\ref{Fig1}(d) and Fig.~\ref{Fig2}(d), it is clear that despite the condition \( r_{\text{p}} / r_{\text{eh}} > 1 \) being satisfied for all values of \( R_0 \), the constraint \( R_{\text{sh}} / r_{\text{p}} > 1 \) is violated for $ R_0<-3.5$. Overall, these figures clearly illustrate that while the electric charge \( q \) and the modified gravity's parameter (encoded in \( f_{R_0} \)) tend to broaden the parameter space for physically consistent shadows, certain curvature terms can impose tighter constraints. This emphasizes the need for a careful balance between gravity and electromagnetic contributions when analyzing shadow-forming conditions in \( F(R) \)-$\mathit{EH}$ black holes.

\begin{figure}[!htb]
    \centering

    % Top row
    \subfloat[$a = 0.2$, $R_0 = 1.5$, $f_{R_0} = 2$]{
        \includegraphics[width=0.31\textwidth]{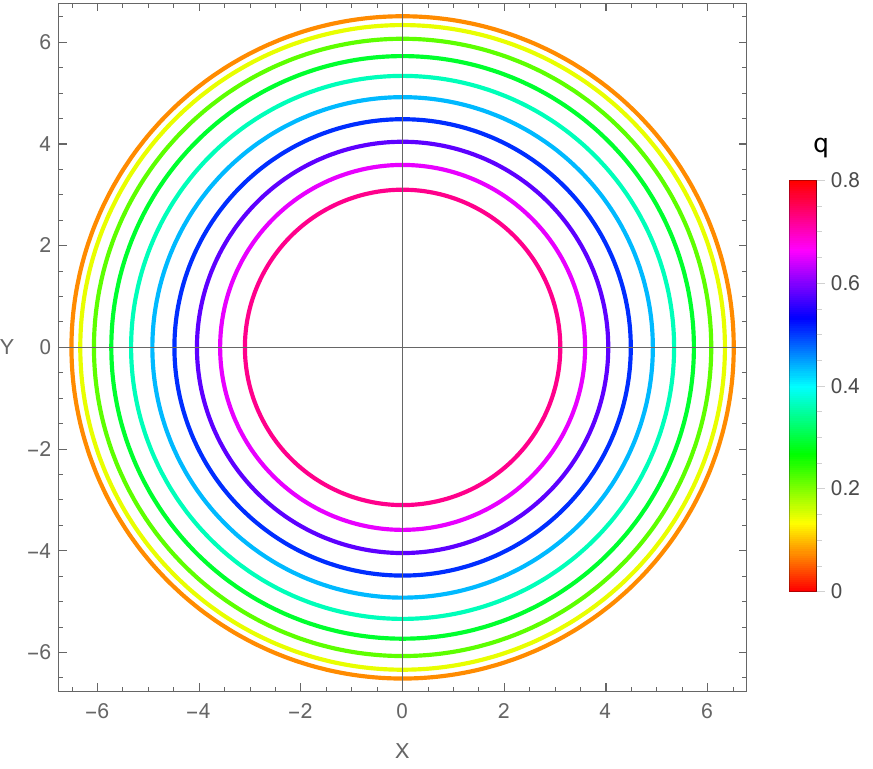}
    }\hfill
    \subfloat[$q = 0.5$, $R_0 = 1.5$, $f_{R_0} = 2$]{
        \includegraphics[width=0.31\textwidth]{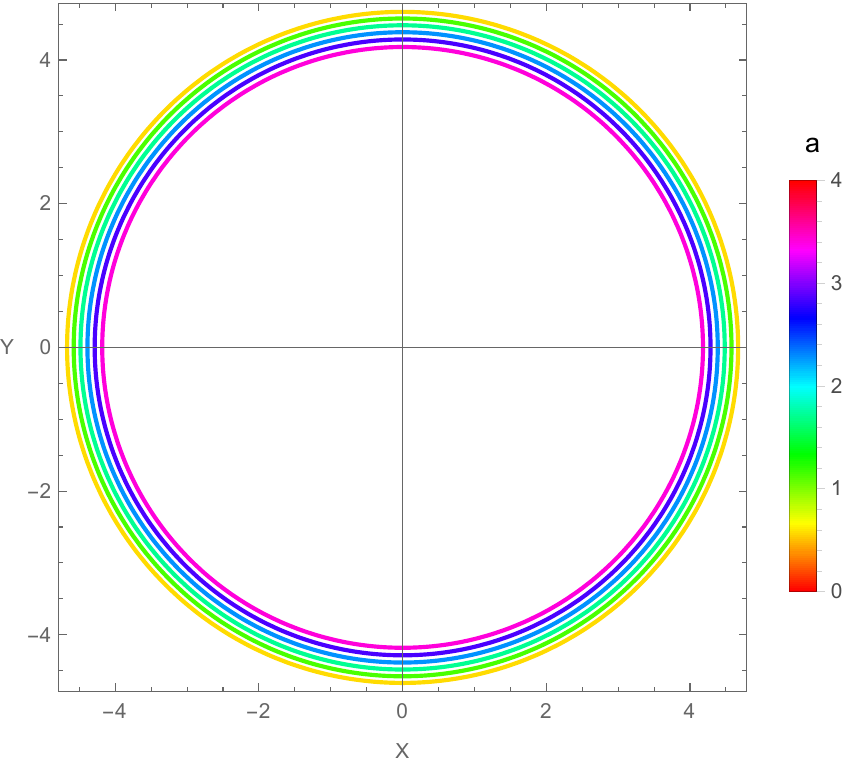}
    }\hfill
    \subfloat[$q = 0.5$, $a = 0.2$, $R_0 = 1.5$]{
        \includegraphics[width=0.31\textwidth]{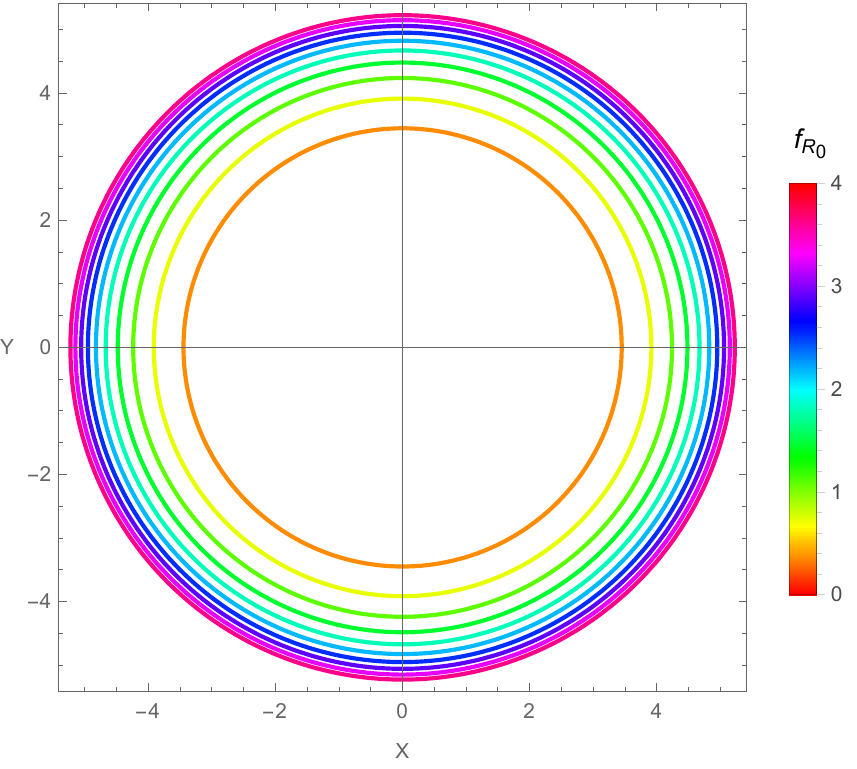}
    }\\[1.5ex]

    % Bottom row
    \subfloat[$q = 0.5$, $a = 0.2$, $f_{R_0} = 2$]{
        \includegraphics[width=0.31\textwidth]{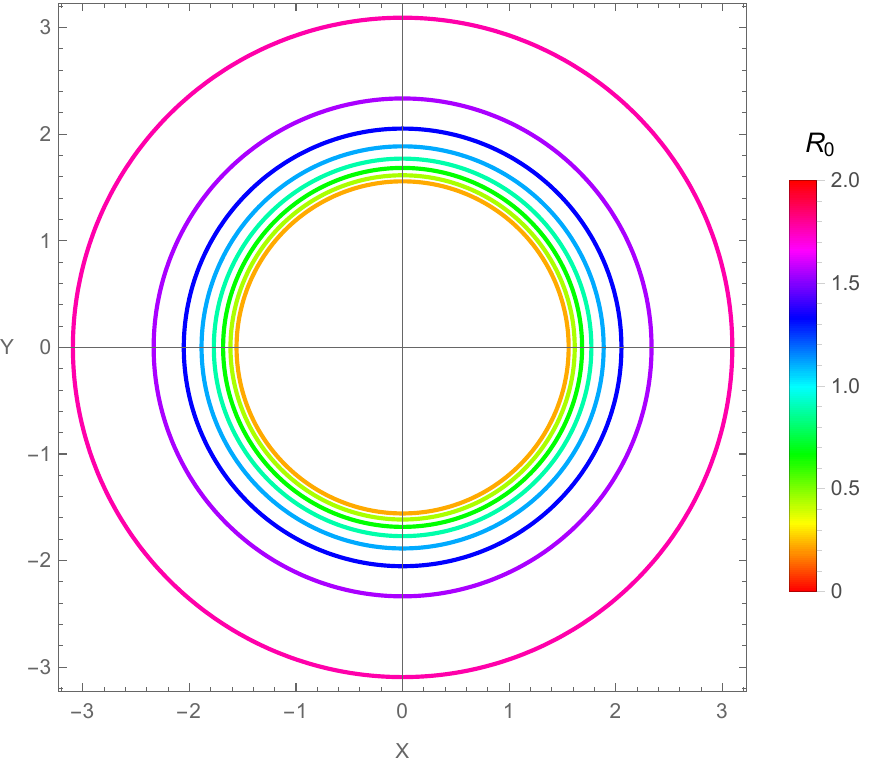}
    }
    \subfloat[$q = 0.5$, $a = 0.2$, $f_{R_0} = 2$]{
        \includegraphics[width=0.31\textwidth]{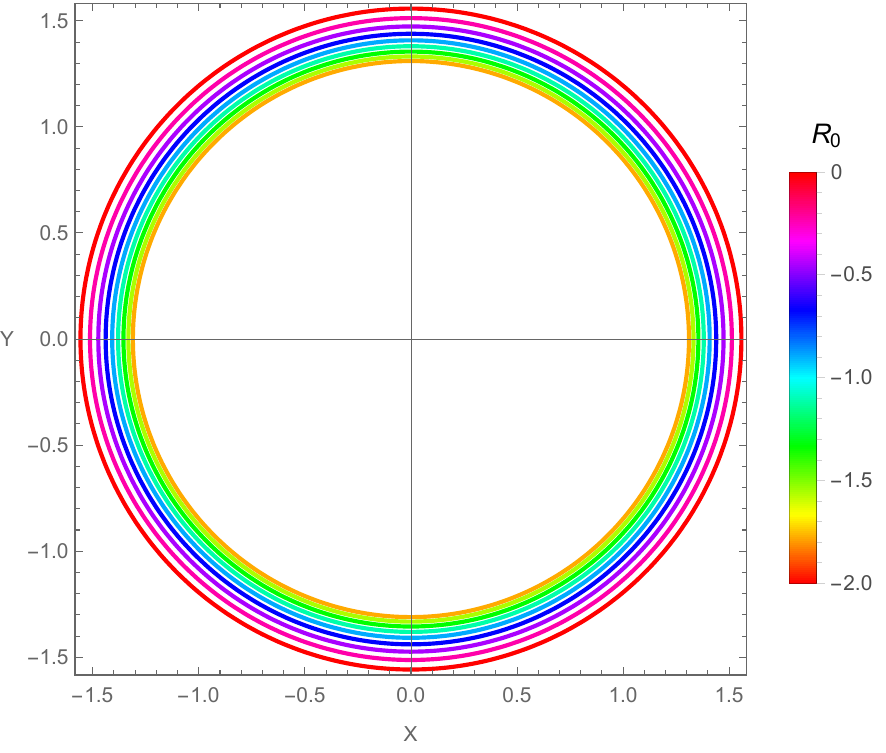}
    }

    \caption{The boundary of the black hole shadow for different values of the parameters $q$, $a$, $f_{R_0}$, and $R_0$.}
    \label{Fig4}
\end{figure}

Fig.~\ref{Fig4} illustrates how the black hole shadow is affected by variations in the physical parameters \( q \), \( a \), \( f_{R_0} \), and \( R_0 \) within the framework of the \( F(R) \)-$\mathit{EH}$ theory. Each subfigure shows the shadow contours for a fixed combination of three parameters, while the fourth is varied. These boundary plots provide insight into how NLED and modified gravity influence the shadow size and shape. Fig.~\ref{Fig4}(a) explores the influence of the electric charge \( q \) for fixed \( a \), \( f_{R_0}  \) and \( R_0  \). It is evident that a higher charge leads to a more contracted shadow, consistent with the repulsive nature of electromagnetic fields. Fig.~\ref{Fig4}(b) shows the impact of the $\mathit{EH}$ nonlinearity parameter \( a \), with \( q \), \( R_0\) and \( f_{R_0}  \) fixed. As \( a \) increases, the shadow shrinks, suggesting that the NLED corrections suppress the photon escape region; however, its effect is negligible. In Fig.~\ref{Fig4}(c), with \( q \), \( a  \) and \( R_0 \) held constant, the effect of increasing \( f_{R_0} \) is examined. The shadow radius visibly increases as \( f_{R_0} \) grows, indicating that stronger modifications to gravity enhances the shadow boundary. Fig.~\ref{Fig4}(d) and \ref{Fig4}(e) analyze the role of curvature corrections through \( R_0 \) in dS and AdS spacetimes, respectively.  It can be seen that in dS spacetime, larger \( R_0 \) values enhance the shadow size under constant \( a \), \( q \), and \( f_{R_0} \), showing how background curvature influences photon orbits. While in AdS spacetime, $R_{0}$ effect will be the opposite.

These results demonstrate that the shadow boundary is sensitive to both gravitational and electromagnetic parameters, making black hole shadow observations a powerful tool to probe the underlying theory.
These analyses are essential for identifying the physically consistent regions of the parameter space and for understanding how NLED and modifications to gravity, such as those introduced in \( F(R) \)-$\mathit{EH}$ theory, affect the observable properties of black holes, particularly their shadows.

\subsection{Observational constraints from the EHT}\label{sec4c}
The study of black hole shadows has become particularly significant with the advent of high-resolution observations, such as those provided by the EHT \cite{akiyama2019event,akiyama2019even}, which captured the first image of the shadow of a supermassive black hole M87*. Comparing theoretical predictions of black hole shadows with observational data provides a powerful means of testing the validity of modified gravity models and evaluating how closely each model matches the observed phenomena.
%The {distortion} of the black hole shadow refers to the deviation from a perfectly circular shape. It is relevant only for rotating black holes since static black holes produce circular shadows by definition. Therefore, shadow distortion is typically studied when comparing rotating black holes to static ones, as well as to identify black holes with the least distortion.The distortion is quantified using an observable known as the radius of the shadow \( R_{\text{sh}} \), which is given in (\ref{equcg}). 

\begin{figure}[!htb]
    \centering
    \subfloat[$a = 0.5$, $f_{R_0} = 2$]{
        \includegraphics[width=0.4\textwidth]{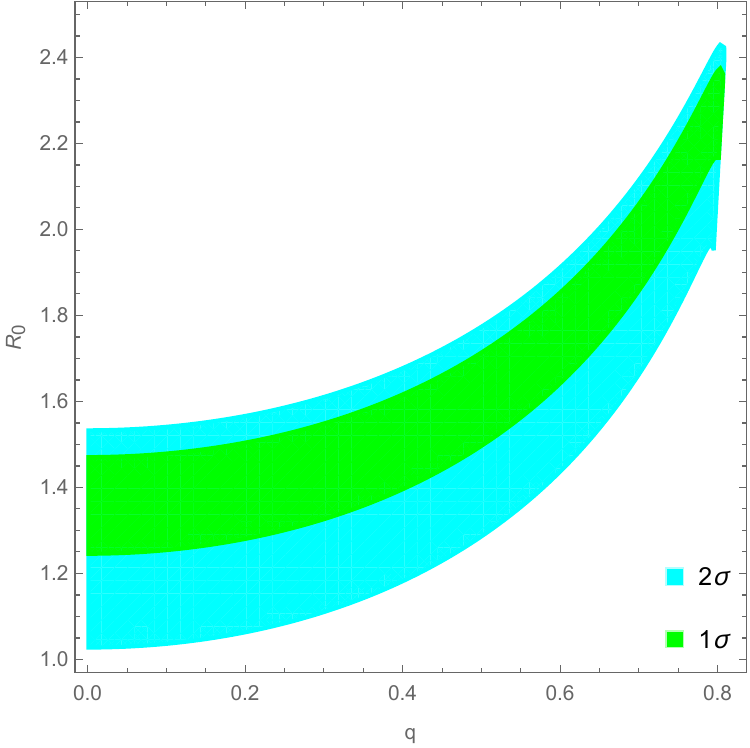}
    }
    \subfloat[$R_0 = 1.5$, $f_{R_0} = 2$]{
        \includegraphics[width=0.4\textwidth]{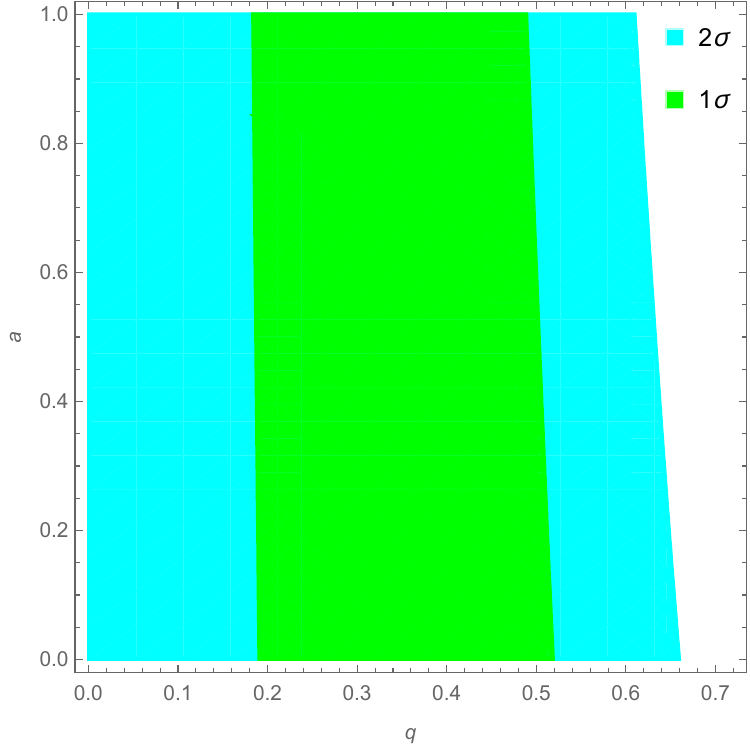}
    }\\[1.5ex]
    \subfloat[$a = 0.5$, $R_0 = 1.5$]{
        \includegraphics[width=0.4\textwidth]{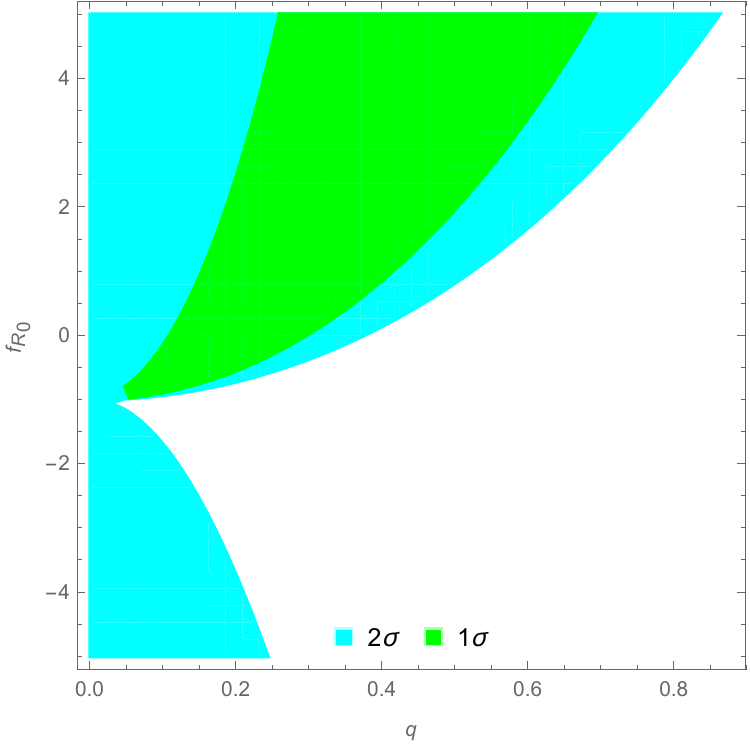}
    }
    \subfloat[$q = 0.2$, $a = 0.5$]{
        \includegraphics[width=0.4\textwidth]{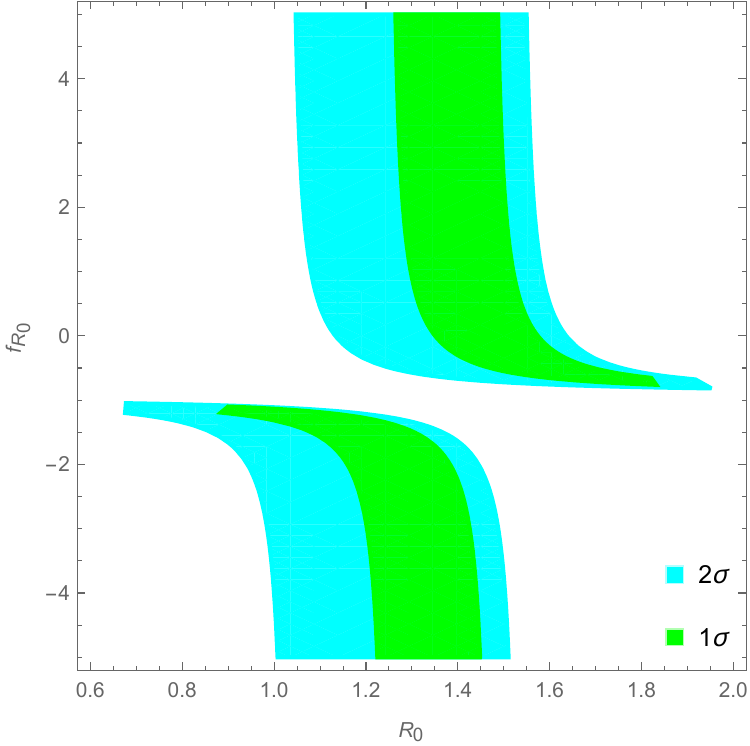}
    }
    \caption{Constraints on parameters of the F(R)-EH-dS black hole based on the EHT observations of $M87^*$.}
    \label{FigEHT1}
\end{figure}
In this study, we compare our theoretical shadow predictions with the shadow observations of M87*  provided by the EHT \cite{zahid2023shadow}. 
%It is evident that the angular diameter depends not only on the black hole's parameters and the observer's inclination angle but also implicitly on the mass of the black hole.
For M87*, we adopt a mass of \( M = 6.5 \times 10^9 M_\odot \) and a distance of \( d = 16.8 \, \text{Mpc} \) from Earth \cite{akiyama2019event}. For simplicity, we do not incorporate uncertainties in mass or distance in this analysis. According to EHT measurements, the angular diameter of the M87* shadow is \( \theta_d = 42 \pm 3 \, \mu\text{as} \) at the 1-\( \sigma \) confidence level \cite{akiyama2019even,yasmin2025shadow}. 
 Using the angular diameter of the black hole shadow and the distance to M87* from Earth, the shadow diameter in units of mass is calculated as
\begin{align} \label{equdtheta} 
D_{M87*} = \frac{d_\theta }{M} \approx 11.0 \pm 1.5
\end{align}
From  (\ref{equdtheta}), the value of \( D_{M87^{*}} \) lies within 1-$\sigma$ confidence as \( 9.5 \leq D_{M87^{*}} \leq 12.5 \), and within 2-$\sigma$ confidence, it falls in the range \( 8 \leq D_{M87^{*}} \leq 14 \).

Using the computed shadow diameter \( d_{sh} \) and deviation parameter \( \delta \), we have constrained the black hole parameters in the $F(R)-EH$ framework using EHT data from M87*. Table~\ref{tab:M87_constraints} lists the allowed intervals of the model parameters \( q \), \( R_0 \), and  \( a \). For fixed values of \( a \) and \( f_{R_0} \), we identified ranges of the charge-like parameter \( q \) and the curvature scalar \( R_0 \) that yield shadow diameters consistent with the EHT observational bounds at 1$\sigma$ and 2$\sigma$ confidence levels. The results demonstrate tighter constraints in the 1$\sigma$ region, while broader values are permitted within 2$\sigma$, emphasizing the sensitivity of the shadow profile to nonlinear electrodynamics and curvature corrections in this modified gravity context. 
\begin{figure}[!htb]
    \centering
    \subfloat[$a = 0.5$, $f_{R_0} = -1.1$]{
        \includegraphics[width=0.4\textwidth]{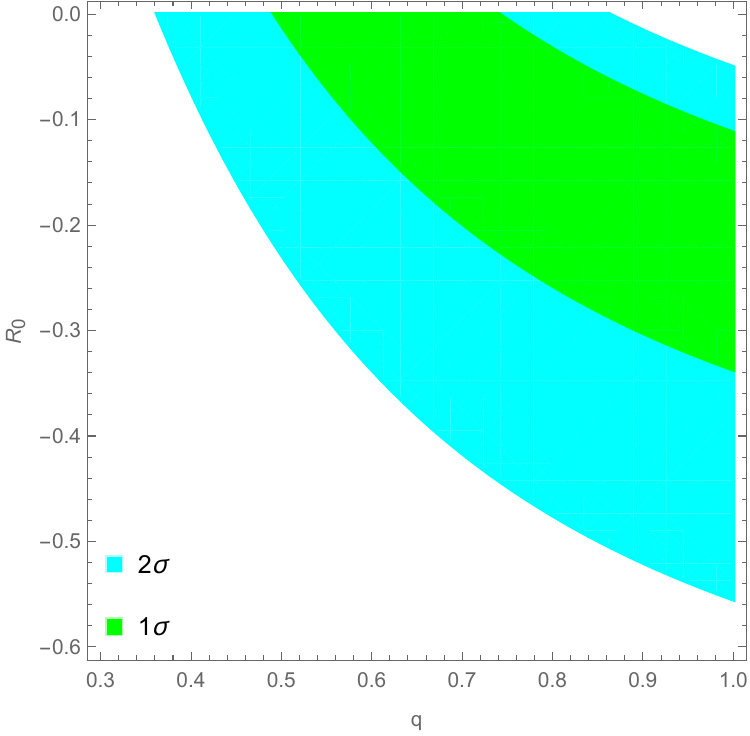}
    }
    \subfloat[$R_0 = -0.1$, $f_{R_0} = -1.1$]{
        \includegraphics[width=0.4\textwidth]{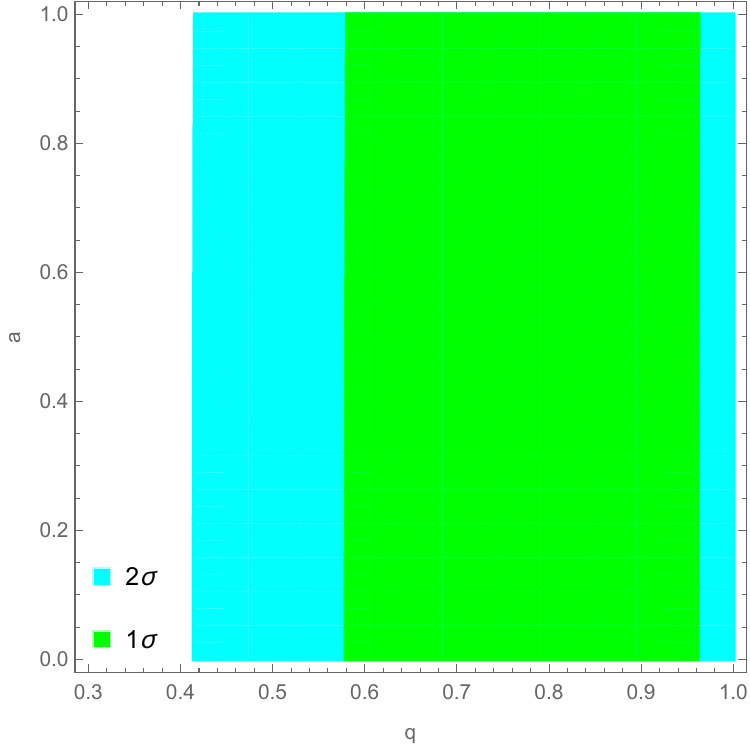}
    }\\[1.5ex]
    \subfloat[$a = 0.5$, $R_0 = -0.1$]{
        \includegraphics[width=0.4\textwidth]{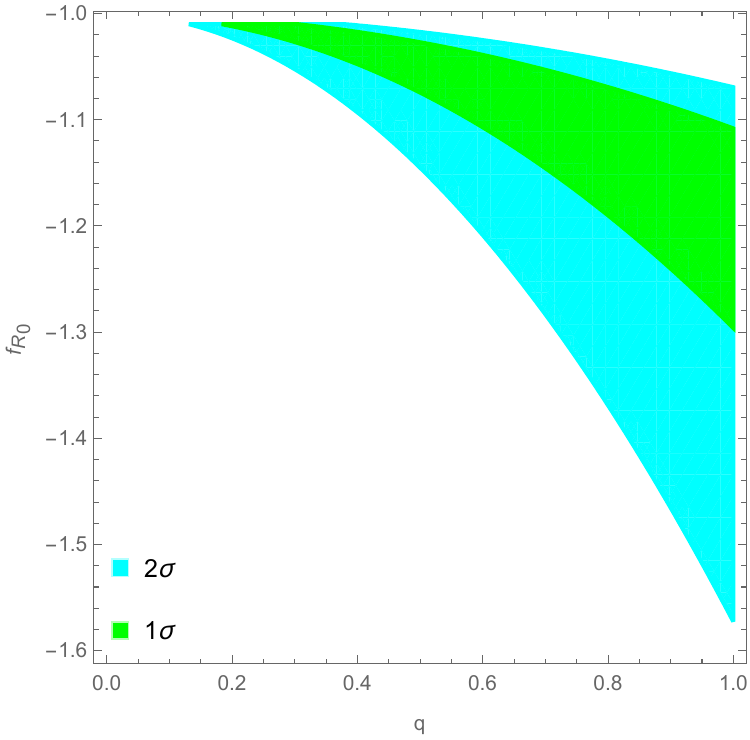}
    }
    \subfloat[$q = 0.9$, $a = 0.5$]{
        \includegraphics[width=0.4\textwidth]{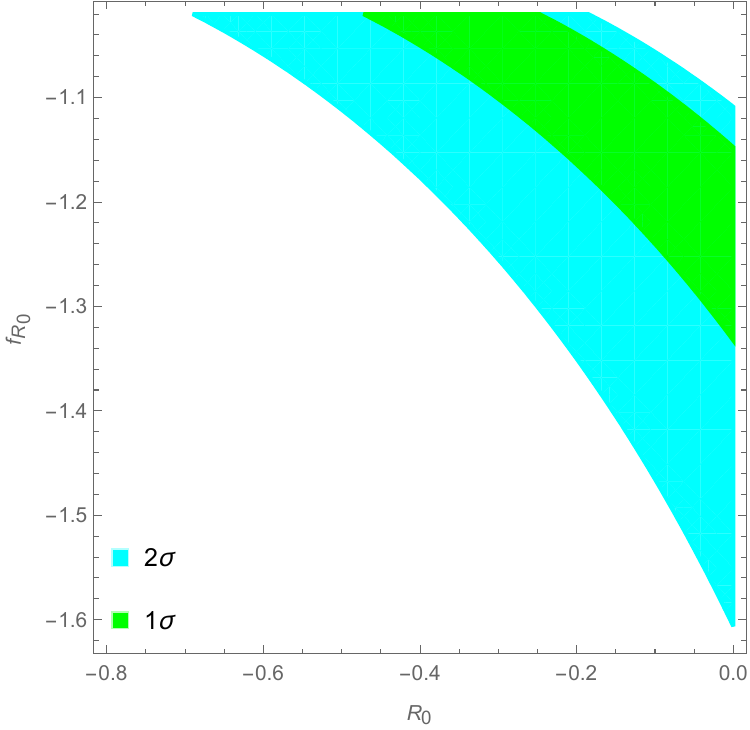}
    }
    \caption{Constraints on parameters of the F(R)-EH-AdS black hole based on the EHT observations of $M87^*$.}
    \label{FigEHT2}
\end{figure}
\begin{figure}[!htb]
    \centering
    \subfloat[$a = 0.5$, $f_{R_0} = 2$]{
        \includegraphics[width=0.4\textwidth]{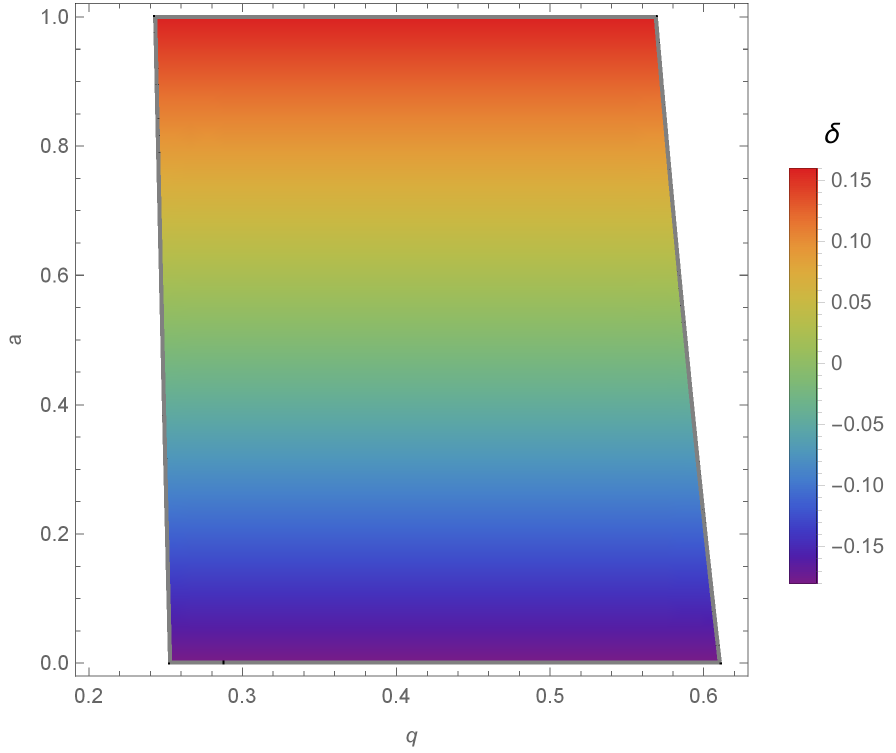}
    }
    \subfloat[$R_0 = 1.5$, $f_{R_0} = 2$]{
        \includegraphics[width=0.4\textwidth]{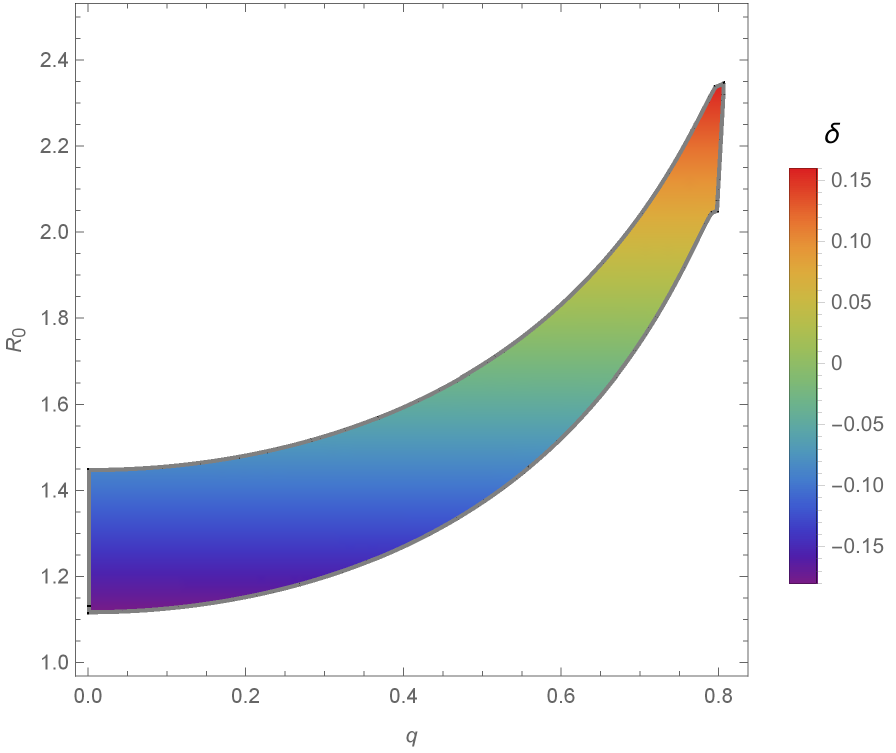}
    }\\[1.5ex]

    % Bottom row
    \subfloat[$a = 0.5$, $R_0 = 1.5$]{
        \includegraphics[width=0.4\textwidth]{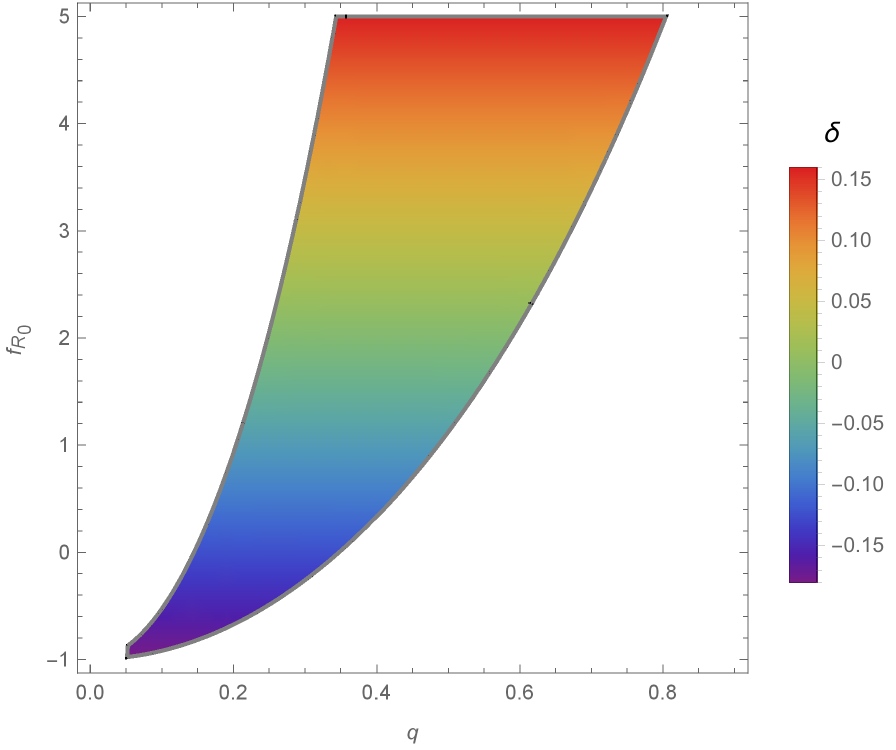}
    }
    \subfloat[$q = 0.2$, $a = 0.5$]{
        \includegraphics[width=0.4\textwidth]{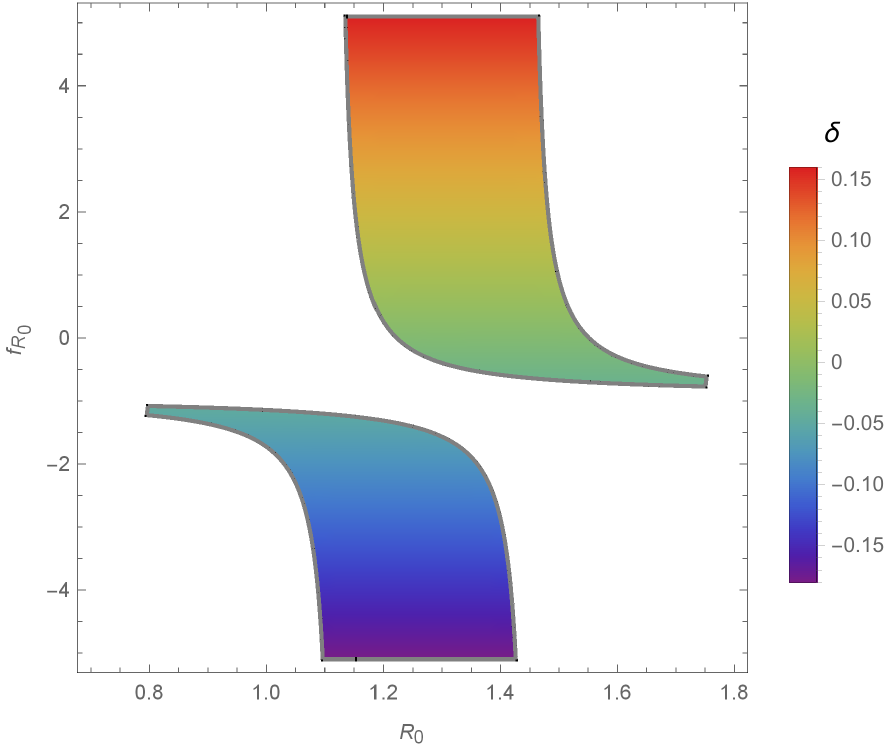}
    }
    \caption{The shadow diameter deviation of the F(R)-EH-dS black hole from a Schwarzschild black hole as a function of model parameters. Note that the colorless regions are forbidden for $q$, $a$, $R_0$, and $f_{R_0}$.}
    \label{Figdelta1}
\end{figure}
To determine the allowed regions of the model parameters, for which the constraint \eqref{equdtheta} is satisfied, we present Fig.~ \ref{FigEHT1} for F(r)-EH BHs in dS spacetime, and Fig.~ \ref{FigEHT2} for those in AdS spacetime. In Figs.~ \ref{FigEHT1} and~ \ref{FigEHT2}, the green-shaded areas indicate the regions that are consistent with observational data within the 1$\sigma$ confidence level, while the cyan-shaded areas represent the regions compatible with the EHT data within the 2$\sigma$ uncertainty range. In our analysis, we fix two of the four model parameters and explore the allowed parameter space for the remaining two. Our results indicate that the shadow of AdS black holes cannot be reconciled with observational data when the electric charge is very small and 
$f_{R_0} > -1$. In contrast, this constraint does not apply to dS black holes. This suggests that the F(R)-EH-dS black hole is a viable candidate for explaining the supermassive black hole M87*.The EHT collaboration on the supermassive black hole M87* introduced an important observable known as the {Schwarzschild shadow deviation}, denoted by \( \delta \). This quantity measures the relative difference between the predicted shadow diameter from a given model (\( d_{\text{metric}} \)) and the shadow diameter of a Schwarzschild black hole. It is defined as \cite{akiyama2022first}
\begin{equation}
\delta = \frac{d_{\text{metric}}}{6\sqrt{3}} - 1
\end{equation}
Here, \( d_{\text{metric}} = 2 R_{\text{sh}} \) represents the model-predicted diameter of the black hole shadow. Based on the EHT observations of M87*, the deviation from the Schwarzschild black hole shadow is constrained within the range \(
-0.18 < \delta < 0.16
\).
\begin{table}[h]
\centering
    \caption{\label{tab1}{Constraints on BH parameters using the EHT data of M87*.}}
\resizebox{\textwidth}{!}{%
\begin{tabular}{|c|c|c|c|c|}
\hline 
\multirow{3}{*}{Observable} & \multicolumn{2}{|c|}{F(R)-EH-dS} & \multicolumn{2}{|c|}{F(R)-EH-AdS} \\ 
\cline{2-5}
& \multicolumn{2}{|c|}{$a = 0.5,\ R_0=1.5,\ f_{{R}_0} = 2$} & \multicolumn{2}{|c|}{$a = 0.5,\ R_0=-0.1,\ f_{{R}_{0}} = -1.1$} \\ 
\cline{2-5}
& $1 \sigma$ & $2 \sigma$ & $1 \sigma$ & $2 \sigma$ \\ 
\hline
$d_{sh}$ & $q \in [0.187,\ 0.502)$ & $q \in [0.0, \ 0.187)$ & $q \in (0.579,\ 0.962]$ & $q \in (0.416,\ 0.579]$ \\ 
 &  & $ ~~q \in [0.502,\ 0.631]$ &  &  \\
\hline
$\delta$ & $q \in [0.248,\ 0.589)$ & $q \in [0.0, \ 0.248) $ & $q \in (0.471,\ 0.889]$ & $q \in [0.268,\ 0.471]$ \\ 
 &  & $~~~ q \in [0.589,\ 0.741)$ &  &  \\
\hline
\multirow{2}{*}{Observable} & \multicolumn{2}{|c|}{$q = 0.2,\ a=0.5,\ f_{{R_0}} = 2$} & \multicolumn{2}{|c|}{$q = 0.9, \ a=0.5,\ f_{{R_0}} = -1.1$} \\ 
\cline{2-5}
& $1 \sigma$ & $2 \sigma$ & $1 \sigma$ & $2 \sigma$ \\ 
\hline
$d_{sh}$ & $R_0 \in [1.281, \ 1.504]$ & $R_0 \in [1.065,\ 1.281)$ & $R_0 \in [-0.302,\ -0.078]$ & $R_0 \in (-0.078,\ -0.015)$ \\
&  & $ R_0 \in [1.504, \ 1.566]$ & & $ R_0 \in [-0.521,\ -0.302)$ \\
\hline
$\delta$ & $R_0 \in [1.151,\ 1.481]$ & $R_0 \in [0.729, \ 1.151)$ & $R_0 \in [-0.432,\ -0.101]$ & $R_0 \in [-0.856,\ -0.432) $ \\ 
 &  & $ R_0 \in [1.481,\ 1.556]$ &  & $ R_0 \in [-0.101,\ -0.026)$ \\ 
\hline
\multirow{2}{*}{Observable} & \multicolumn{2}{|c|}{$q = 0.2,\ R_0 = 1.5, \ f_{{R_0}} = 2$} & \multicolumn{2}{|c|}{$q = 0.9,\ R_0 = -0.1, \ f_{{R_0}} = -1.1$} \\ 
\cline{2-5}
& $1 \sigma$ & $2 \sigma$ & $1 \sigma$ & $2 \sigma$ \\ 
\hline
$d_{sh}$ & $a \in [0, \ 1]$ & $---$ & $a \in [0,\ 1]$ & $---$ \\ 
\hline
$\delta$ & $--$ & $a \in [0,\ 1]$ & $---$ & $a \in [0,\ 1]$ \\ 
\hline
\multirow{2}{*}{Observable} & \multicolumn{2}{|c|}{$q = 0.2, \ a = 0.5,\ R_0 = 1.5$} & \multicolumn{2}{|c|}{$q = 0.9, \ a=0.5,\ R_0 = -0.1$} \\ 
\cline{2-5}
& $1 \sigma$ & $2 \sigma$ & $1 \sigma$ & $2 \sigma$ \\ 
\hline
$d_{sh}$ & $f_{{R_0}} \in [-0.549,\ 2.443]$ & $f_{{R_0}} \in (-0.721, -0.549) $ & $f_{{R_0}} \in [-1.240,\ -1.087)$ & $f_{{R_0}} \in (-1.465,\ -1.240)$ \\ 
 &  & $f_{{R_0}} >2.443$ &  &  \\ 
\hline
$\delta$ & $f_{{R_0}} \in [ -0.676, \ 0.927]$ & $f_{{R_0}} \in [-0.802, \-0.676 )$ & $f_{{R_0}} \in [-1.362, \ -1.101]$ & $f_{{R_0}} \in (-1.101, \ -1.061)$ \\
 & & $ f_{{R_0}} > 0.927$ &  & $f_{{R_0}} \in (-2.072, \ -1.362)$ \\ 
\hline
\end{tabular}%
}
\label{tab:M87_constraints}
\end{table}
Clearly, if the observed black hole shadow is larger than that of a Schwarzschild black hole with the same mass, the deviation parameter \( \delta \) will be positive. Conversely, if the shadow is smaller, \( \delta \) will take on a negative value.
Figs.~ \ref{Figdelta1} and \ref{Figdelta2} show the parameter regions that satisfy the constraint on the shadow diameter \( \delta \), for dS and AdS black holes, respectively. In Fig.~ \ref{Figdelta1}(a), we fix $a$ and $f_{R_0}$ and examine the admissible parameter space in the $(q,R_0)$ plan. It is evident that the shadow of black holes situated in backgrounds with higher curvature and stronger electric fields is larger than that of the Schwarzschild black hole. Fig.~ \ref{Figdelta1}(b) displays the admissible parameter space in the $(q,a)$ plan for fixed $R_0$ and $f_{R_0}$, illustrating that the black hole shadow becomes larger than that of the Schwarzschild black hole for large values of the EH parameter. An interesting result emerges in Fig.~\ref{Figdelta1}(d), which shows the admissible parameter space in the 
$(R_0,f_{R_0})$ plane with fixed $a$ and $q$. When both the electric and NLED fields are weak, the black hole shadow exceeds that of the Schwarzschild black hole for 
$f_{R_0}>-1$, whereas for $f_{R_0}<-1$, it becomes smaller.

Fig.~\ref{Figdelta2}(a) shows that, similar to the dS case, the BH shadow in AdS spacetime becomes larger than that of the Schwarzschild BH when the NLED field is strong. However, in contrast to the dS case, the shadow in AdS spacetime becomes smaller than the Schwarzschild BH shadow for black holes situated in a high-curvature background with a strong electric field. As previously mentioned, for AdS BHs, consistency with the EHT data is observed when $f_{R_{0}}<-1$. 
Fig.~\ref{Figdelta2}(c) shows that only for 
$f_{R_{0}}$ values close to $ -1 $, the BH shadow becomes larger than that of the Schwarzschild BH.
A closer examination of Fig.~\ref{Figdelta2} reveals that, for the small values of modified gravity parameters \( |R_0| \) and \( |f_{R_0}| \), an AdS black hole in the F(R)-$\mathit{EH}$ framework with the same mass as a Schwarzschild black hole produces a larger shadow. In contrast, for charged dS black holes, the shadow size exceeds that of the Schwarzschild black hole only when the parameters take on large values.
\begin{figure}[!htb]
    \centering
    \subfloat[$a = 0.5$, $f_{R_0} = -1.1$]{
        \includegraphics[width=0.4\textwidth]{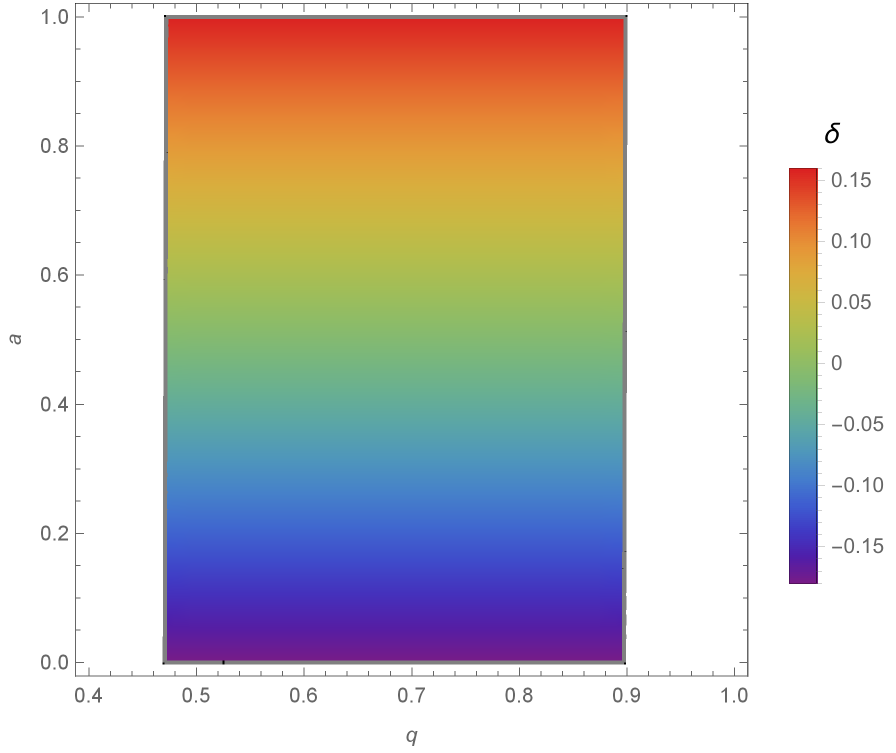}
    }
    \subfloat[$R_0 = -0.1$, $f_{R_0} = -1.1$]{
        \includegraphics[width=0.4\textwidth]{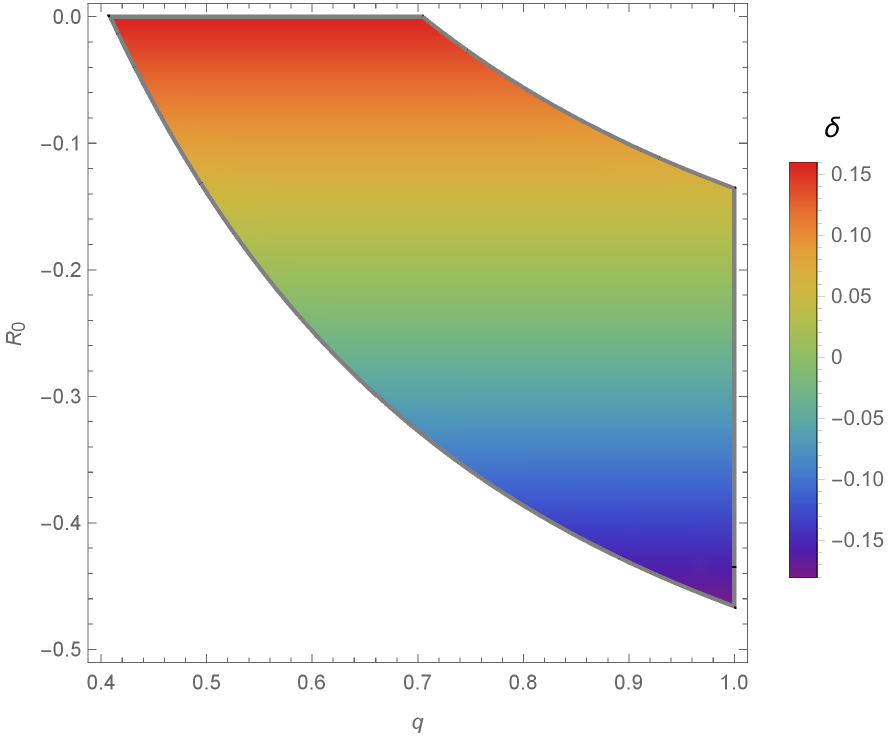}
    }\\[1.5ex]

    % Bottom row
    \subfloat[$a = 0.5$, $R_0 = -0.1$]{
        \includegraphics[width=0.4\textwidth]{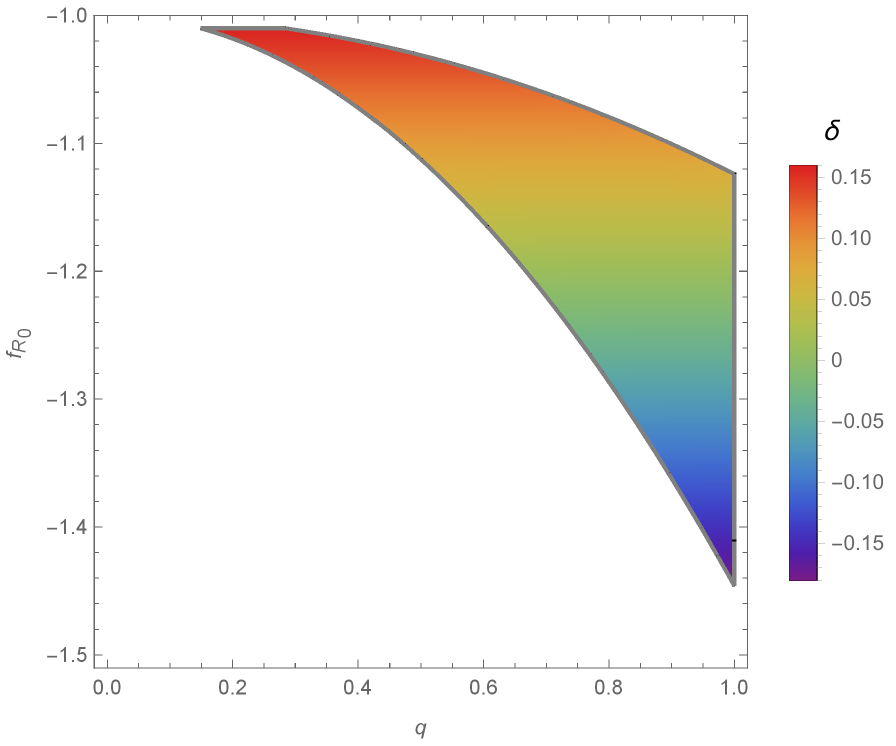}
    }
    \subfloat[$q = 0.2$, $a = 0.5$]{
        \includegraphics[width=0.4\textwidth]{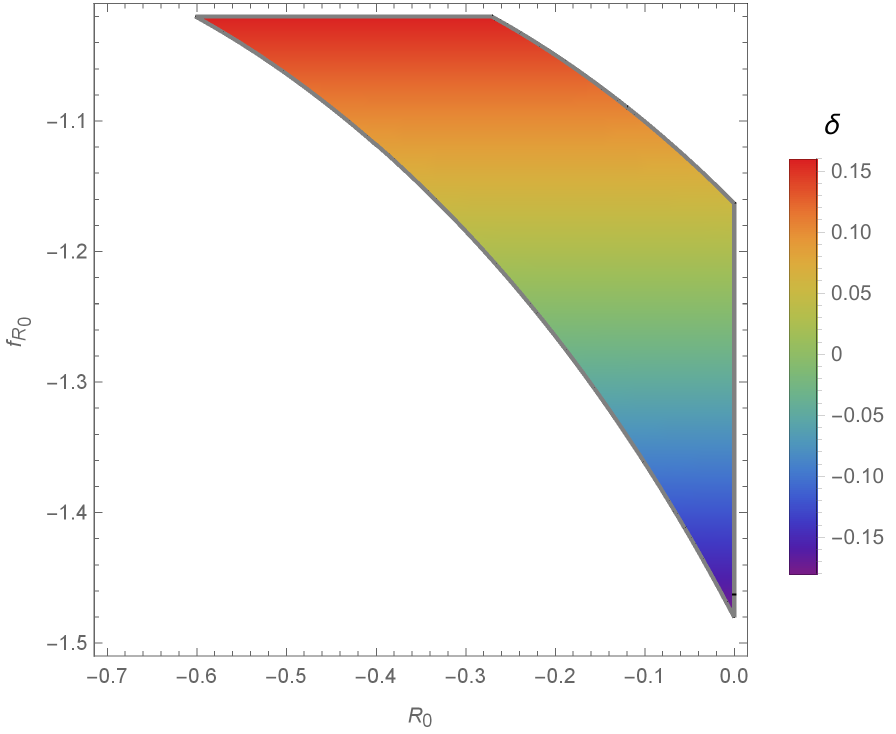}
    }
    \caption{The shadow diameter deviation of the F(R)-EH-dS black hole from a Schwarzschild black hole as a function of model parameters. Note that the colorless regions are forbidden for $q$, $a$, $R_0$, and $f_{R_0}$.}
    \label{Figdelta2}
\end{figure}
\subsection{Energy emission rate} \label{sec4d}
In classical GR, any object falling into a black hole is lost forever. However, quantum mechanically, black holes are believed to emit radiation due to quantum fluctuations near the event horizon. These fluctuations can result in the creation of particle-antiparticle pairs, with one particle escaping the black hole's gravitational pull via quantum tunneling, a process that leads to black hole evaporation.
 The radiation escaping is typically described by the energy emission rate, which is connected to the black hole shadow in the high energy limit. In this regime, the absorption cross-section is approximately equal to the area of the shadow, and is given by \cite{decanini2011u}
\begin{equation}
\sigma_{\text{lim}} \approx \pi R_{\text{sh}}^2.
\end{equation}
The energy emission rate per unit frequency per unit time is expressed as \cite{wei2013observing}
\begin{equation}
\frac{d^2 \mathcal{E}(\omega)}{d\omega \, dt} = \frac{2\pi^2 \sigma_{\text{lim}} \omega^3}{e^{\omega / T_H} - 1}
\approx
\frac{2\pi^3 R_{\text{sh}}^2 \omega^3}{e^{\omega / T_H} - 1}.
\end{equation}
Here, \( \omega \) is the angular frequency of the emitted radiation, and \( T_H \) is the Hawking temperature, given by
\begin{equation}
T_H = \frac{1}{r_{eh}^2}- \frac{R_0}{6}r_{eh}+\frac{1}{1+f_{R_0}}\left(\frac{3 a q^4}{10 r_{eh}^7}-\frac{2q^2}{r_{eh}^3}\right).
\end{equation}
These results allow us to estimate the Hawking temperature and radiation spectrum of electrically charged black holes in \( F(R) \)-$\mathit{EH}$ theory. The shadow radius \( R_{\text{sh}} \), determined from earlier sections, directly affects the total energy emission, providing a bridge between black hole thermodynamics and observational features such as the shadow profile. 
\begin{figure}[!htb]
		\centering
		\subfloat[  $ a=0.2 $, $ R_{0}=1.5 $, $f_{R_{0}}=2 $]{
			\includegraphics[width=0.31\textwidth]{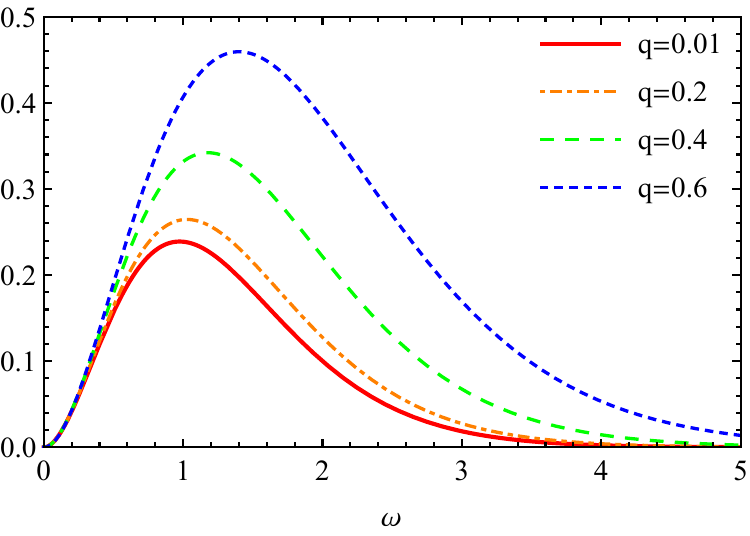}}
		\subfloat[ $ q=0.5 $, $ R_{0}=1.5 $, $f_{R_{0}}=2 $]{
			\includegraphics[width=0.31\textwidth]{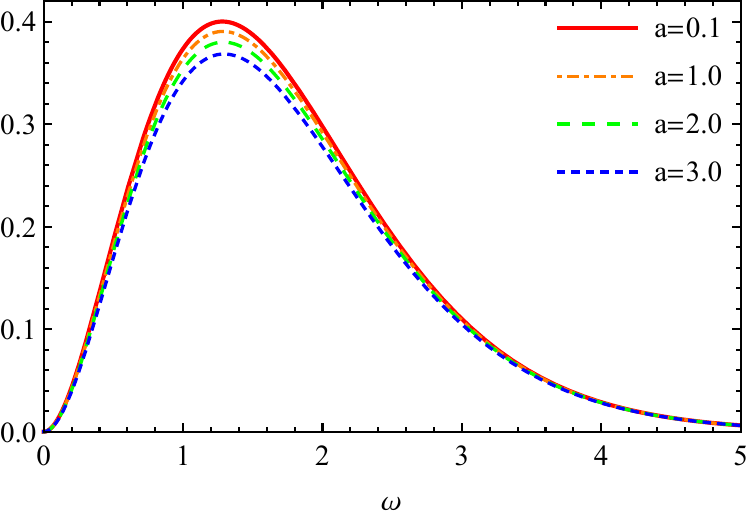}}
			\subfloat[$ q=0.5 $, $ a=0.2 $, $ R_{0}=1.5 $]{
			\includegraphics[width=0.31\textwidth]{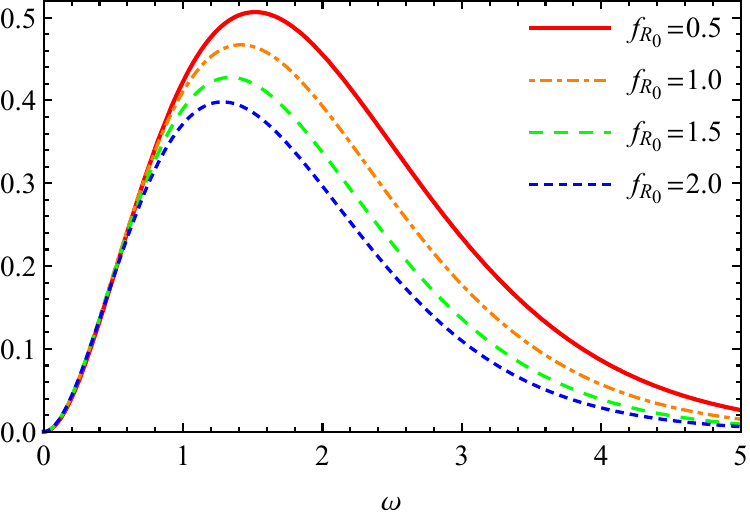}}       
		\newline
		\subfloat[$ q=0.5 $, $ a=0.2 $, $f_{R_{0}}=2 $]{
			\includegraphics[width=0.31\textwidth]{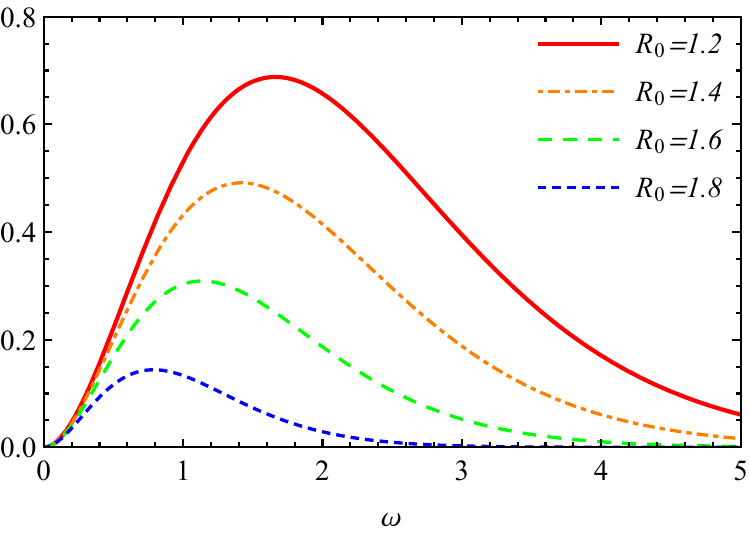}}
		\subfloat[$ q=0.5 $, $ a=0.2 $, $f_{R_{0}}=2 $]{
			\includegraphics[width=0.31\textwidth]{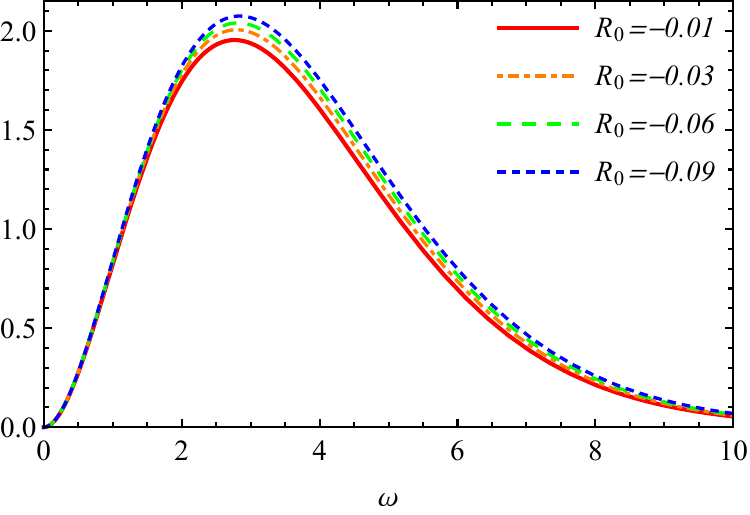}}        
		\newline
		\caption{The energy emission rates for the F(R)-$\mathit{EH}$ black holes with different values of parameters $ q $, $ a $, $ f_{R_{0}}$ and $R_{0}$.}
		\label{FigEr}
	\end{figure}
Fig.~\ref{FigEr} presents the behavior of the energy emission rate as a function of the emission frequency for various values of the parameters \( q \), \( a \), \( f_{R_0} \), and \( R_0 \) in our modified \( F(R) \)-based spacetime. It is evident that the energy emission curve exhibits a peak that shifts depending on the chosen parameters.
From Fig.~\ref{FigEr}(a), we observe that increasing the electric charge \( q \) enhances the peak energy emission rate. This implies that charged black holes emit radiation more rapidly than neutral ones, resulting in a faster evaporation process and consequently shorter lifetimes. Fig.~\ref{FigEr}(b) shows that an increase in $\mathit{EH}$ parameter $ a $ leads to a reduction in the emission rate. This suggests that stronger NLED effects slow down the black hole evaporation process, resulting in longer lifetimes. The influence of the gravitational modification parameter 
\( f_{R_0} \) is illustrated in Fig.~\ref{FigEr}(c). An increase in \( f_{R_0} \) leads to a decrease in the emission rate, implying that black holes in 
\( F(R) \) gravity evaporates more slowly than that in GR. Figs.~\ref{FigEr}(d) and \ref{FigEr}(e) demonstrate the role of the curvature background, governed by \( R_0 \). In asymptotically dS spacetime, higher curvature values are associated with longer lifetimes. In contrast, in AdS spacetime, a higher curvature background accelerates black hole evaporation, leading to shorter lifetimes.
    \section{Summary and Discussions} \label{sec5}
    \begin{justify}
In this work, we analyzed the shadow and energy emission properties of a static, spherically symmetric electrically charged black hole within the \( F(R) \) gravity framework coupled with $\mathit{EH}$ NLED. Our analysis reveals that the electric charge \( q \) and the modified gravity parameter \( f_{R_0} \) contribute to an expansion of the photon ring and lensing regions, whereas the parameter \( R_0 \) has a suppressing effect, leading to a reduction in the extent of these regions. In contrast to the $\mathit{EH}$ parameter \( a \), which exhibits a more limited influence, the three parameters \( q \), \( f_{R_0} \), and \( R_0 \) play distinct and significant roles in determining photon trajectories.

The black hole shadow in \( F(R) \)$-\mathit{EH}$ theory is defined by the photon sphere radius \( r_{\text{p}} \), the event horizon radius \( r_{\text{eh}} \), and the shadow radius \( R_{\text{sh}} \), which satisfy the condition
$r_{\text{eh}} < r_{\text{p}} < R_{\text{sh}}$.
This ensures a physically viable and observable shadow structure. Our analysis identifies admissible parameter regions where this condition holds, corresponding to physically consistent black hole solutions. 
We find that increasing the electric charge \( q \) and the modified gravity parameter \( f_{R_0} \) expands these admissible regions, allowing a wider range of parameters that produce well-defined shadows. The electric charge tends to shrink the shadow radius due to electromagnetic repulsion, while the $\mathit{EH}$ parameter \( a \) has a subtler shrinking effect. Conversely, stronger gravity modifications (larger \( f_{R_0} \)) enlarge the shadow. Regarding curvature effects, our findings show that increasing \( R_0 \) enhances the shadow size in dS spacetime, whereas increasing \( \vert R_0 \vert \) decreases the shadow radius in AdS backgrounds. These results highlight the crucial interplay between electromagnetic and gravitational parameters in shaping black hole shadows, making them effective observational probes for testing alternative gravity theories.

Using the observational constraints for M87* detailed previously, we constrain the parameter space of \( F(R) \)-$\mathit{EH}$ black holes in both dS and AdS spacetimes. Parameter regions consistent with the EHT shadow measurements within 1-\(\sigma\) and 2-\(\sigma\) confidence levels are identified. Notably, AdS black holes with very small electric charge and $f_{R_0} > -1$ fail to satisfy the constraints imposed by EHT data, whereas de Sitter configurations remain fully compatible without any limitations, indicating their viability as models for M87*.

Further analysis shows that increased electric charge \( q \), higher curvature \( R_0 \), and larger NLED parameter \( a \) generally enlarge the shadow size relative to Schwarzschild black holes in dS spacetime. In contrast, in AdS backgrounds, strong curvature combined with a significant electric charge and a large modified gravity parameter \( f_{R_0} \) (with values far from $-1$) can lead to a reduction in the shadow size.
 These findings provide critical insights into the interplay between modified gravity, NLED, and observational constraints.
 
The Hawking temperature and radiation spectrum of electrically charged black holes in \( F(R) \)-$\mathit{EH}$ theory are directly influenced by the shadow radius \( R_{\text{sh}} \), linking black hole thermodynamics with observable shadow characteristics. An increase in electric charge \( q \) raises the peak energy emission rate, leading to faster evaporation and shorter black hole lifetimes compared to neutral cases. In contrast, larger NLED parameter \( a \) lowers the emission rate, indicating that stronger NLED effects slow evaporation and extend lifetimes. Similarly, a higher modified gravity parameter \( f_{R_0} \) reduces the emission rate, suggesting slower evaporation in \( F(R) \) gravity relative to GR. The curvature scalar \( R_0 \) also plays a crucial role: in asymptotically dS spacetimes, increased curvature lengthens black hole lifetimes, whereas in AdS spacetimes, it shortens them by accelerating evaporation.
\end{justify}
\begin{justify}

Our findings show strong alignment with previous studies conducted in modified gravity frameworks coupled to nonlinear electrodynamics. Similar to earlier results, we find that increasing the modified gravity parameter \( f_{R_0} \) expands the parameter space compatible with physical black hole solutions and observational constraints~\cite{Jafarzade2025byr}. Notably, both AdS and dS black holes exhibit shadow behavior that is highly sensitive to curvature: increasing \( R_0 \) enlarges the shadow in de Sitter backgrounds while reducing it in AdS spacetimes, an effect consistently reported across various models~\cite{Jafarzade2025frmodmax}. The shrinking influence of electric charge on the shadow size is also widely supported in the literature, confirming the role of electromagnetic repulsion in tightening photon orbits~\cite{Zhu2019,Zubair2022}.
Compared to related work, our model shows distinct thermodynamic behavior. While some previous studies suggest that stronger nonlinear electrodynamic coupling enhances the energy emission rate~\cite{Jafarzade2025frmodmax}, our results indicate the opposite: the parameter \( a \) suppresses emission, leading to slower evaporation and longer black hole lifetimes. Likewise, although charge is often associated with reduced emission rates in other models~\cite{Jafarzade2025frmodmax}, we find that increased electric charge accelerates evaporation, particularly in dS backgrounds. These differences point to the sensitivity of black hole thermodynamics to the specific form of nonlinear coupling and gravitational modification.
Finally, both our work and existing literature use the EHT data of M87* to constrain model parameters. However, we observe that while other models require \( f_{R_0} < -1 \) for consistency in AdS spacetimes~\cite{Jafarzade2025frmodmax}, our analysis shows that such values fall outside the observationally viable region. This contrast reinforces the importance of detailed shadow analysis in differentiating between viable modified gravity scenarios and highlights the diagnostic power of high-resolution black hole imaging.

\end{justify}

\bibliographystyle{ieeetr}
\bibliography{reference}
\end{document}